\documentclass{aa}  
\makeatletter
\let\linenumbers\relax
\let\runningpagewiselinenumbers\relax

\let\switchlinenumbers\relax

\makeatother
\nolinenumbers
\usepackage{graphicx}
\usepackage{txfonts}
\usepackage{lipsum}
\usepackage{subcaption}         
\usepackage{caption}
\usepackage{lscape}             
\usepackage{placeins}           
\usepackage{hyperref}
                                
\usepackage[usenames,dvipsnames]{color} 
\begin{document}

   \title{GALACTICNUCLEUS: A high angular resolution JHKs imaging survey of the Galactic centre }

   \subtitle{V. Toward the GNS Second Data Release: Methodology, Photometric and Astrometric Performance}

\authorrunning{}
\titlerunning{}

   \author{\'A. Mart\'inez Arranz\inst{1}
     \and
     R. Sch\"odel\inst{1}
     \and
     H. Bouy\inst{2}
     \and
     F. Nogueras Lara\inst{1}
      }

   \institute{Instituto de Astrof\'isica de Andaluc\'ia (CSIC), Glorieta de la Astronomía s/n, 18008 Granada, Spain\\
             \email{amartinez@iaa.csic.es}
            \and Laboratoire d’Astrophysique de Bordeaux (LAB), Universit\'e de Bordeaux, B\^at. B18N, All\'ee Geoffroy Saint-Hilaire CS 50023,
33615 PESSAC CEDEX, France\\ }

   \date{}

 
   \abstract
   {The center of the Milky Way presents a unique environment of fundamental astrophysical interest. However, its extreme crowding and extinction make this region particularly challenging to study. The GALACTICNUCLEUS survey, a high-angular-resolution near-infrared imaging program, was designed to overcome these difficulties. Its first data release provides a powerful resource for exploring the Galactic center and enabling key discoveries in this extreme environment.
}
   {We present the methodology and first results of a second data release of the GALACTICNUCLEUS survey, which incorporates significant improvements in data reduction, calibration, methodology and a second epoch. In particular, we aim to provide deeper photometry, improved astrometry, and high-precision proper motions for the test fields analysed in this study.} 
   {Observations were obtained with VLT/HAWK-I over two epochs separated by approximately seven years for most pointings, and by four to five years for some. The data were acquired using speckle holography, and in the case of the second epoch, a ground-layer adaptive optics system was also employed. We developed a new reduction pipeline with key improvements, including enhanced distortion corrections and jackknife-based error estimation. For the test fields presented in this work, proper motions were derived using two complementary approaches: (i) relative proper motions, aligning epochs within the survey itself, and (ii) absolute proper motions, tied to the Gaia reference frame. Validation was performed on two representative test fields: one in the Galactic bar and one in the crowded nuclear stellar disk, overlapping with the Arches cluster.}
   {For the fields analysed  in this pilot study, the new release achieves photometry $\sim$1 mag deeper and astrometry $\sim$5 times more precise than then first data release. Proper motions reach an accuracy of $\sim$0.5\,mas\,yr$^{-1}$ relative to Gaia, despite being based solely on two, ground-based epochs. Both relative and absolute methods deliver consistent results. In the Arches field, we were able to recover the cluster with mean velocities consistent with previous HST-based studies. Comparisons with previous catalogs confirm the robustness of our methodology.}
   {The second data release of the GALACTICNUCLEUS survey provides, for the test fields analyzed in this study, the most precise ground-based proper motion catalogs of the Galactic center to date. Once the full data set is reduced and the complete catalog released, its wide spatial coverage and high astrometric and photometric accuracy will enable detailed studies of the outer structure of the Nuclear Stellar Disk, the motions of young stars in regions of ongoing star formation, and the identification of new stellar clusters. The expected quality of the final catalogs will also make them well suited for combination with observations from space-based missions, such as JWST and the Roman Space Telescope..
}


   \maketitle

\section{Introduction}

GALACTICNUCLEUS (GNS
hereafter) is a ground-based near-infrared survey of the central
$\sim$0.3\,deg$^{2}$ of the Galactic Center
\citep[GC,][]{Nogueras-Lara:2018pr,Nogueras-Lara:2019yj}. Because of its high angular resolution of $0.2"$ FWHM, which is reached with a combination of short exposures and speckle holography, it is a factor $\sim$10 less confused that the VISTA Variables in the Via Lactea survey \citep[VVV,][]{Minniti:2010fk,Saito:2012fk}, which covers a much larger field, but at seeing limited resolution. Thus, the GNS  reaches several magnitudes deeper than the former, well
below the red clump. 

GNS has been fundamental in addressing several key open questions about the GC. A fundamental discovery enabled by GNS is the early formation and
recent starburst activity of the nuclear stellar disc (NSD) of the Milky Way
\citep{Nogueras-Lara:2020pp}. GNS has further allowed us to study
interstellar extinction towards the GC\citep{Nogueras-Lara:2020qn},
the Milky Way spiral arms towards the GC  \citep{Nogueras-Lara:2021cf},
estimate the distance towards molecular clouds
\citep{Nogueras-Lara:2021dd,Martinez-Arranz:2022wj}, study the
relationship between the nuclear stellar cluster and NSD \citep{Nogueras-Lara:2021lf,Nogueras-Lara:2022by}, find
an age gradient within the nuclear stellar disc \citep{Nogueras-Lara:2023xd}, detect an excess of
young massive stars in the Sagittarius\,B1 HII region
\citep{Nogueras-Lara:2022jz}, and identify the first new bona fide young
star cluster or association detected in the GC since 30\,years
\citep{Martinez-Arranz:2024gm,Martinez-Arranz:2024nr}. \href{https://archive.eso.org/cms/eso-archive-news/first-data-release-from-the-galacticnucleus-survey.html}{GALACTICNUCLEUS images and data products} are publicly available on the ESO Science Archive. The GNS catalogue has also been incorporated into the JWST Guide Star Catalogue, thus improving the pointing accuracy of the space telescope.

To improve the quality of the data products, we developed an enhanced reduction, analysis, and calibration methodology, which we present here as part of the second data release. To enable the computation of proper motions across the survey area, a second imaging epoch of the GNS field was acquired in the $H$ band in 2022. We refer to this second epoch as GNS\,II, and to the earlier data as GNS\,I.

The new data release GNS\,I\,DR2, provides about 1\,mag deeper photometry and five times more accurate absolute astrometry than GNS\,I\,DR1. The proper motion measurements, derived from combining GNS\,I DR2 with GNS\,II, reach an accuracy of 0.5\,mas\,yr$^{-1}$ rms with respect to Gaia data release 3 \citep[GAIA DR3,][]{Gaia-Collaboration:2024sw} . 

\begin{figure*}[!h]
		\centering
		\includegraphics[width=\linewidth]{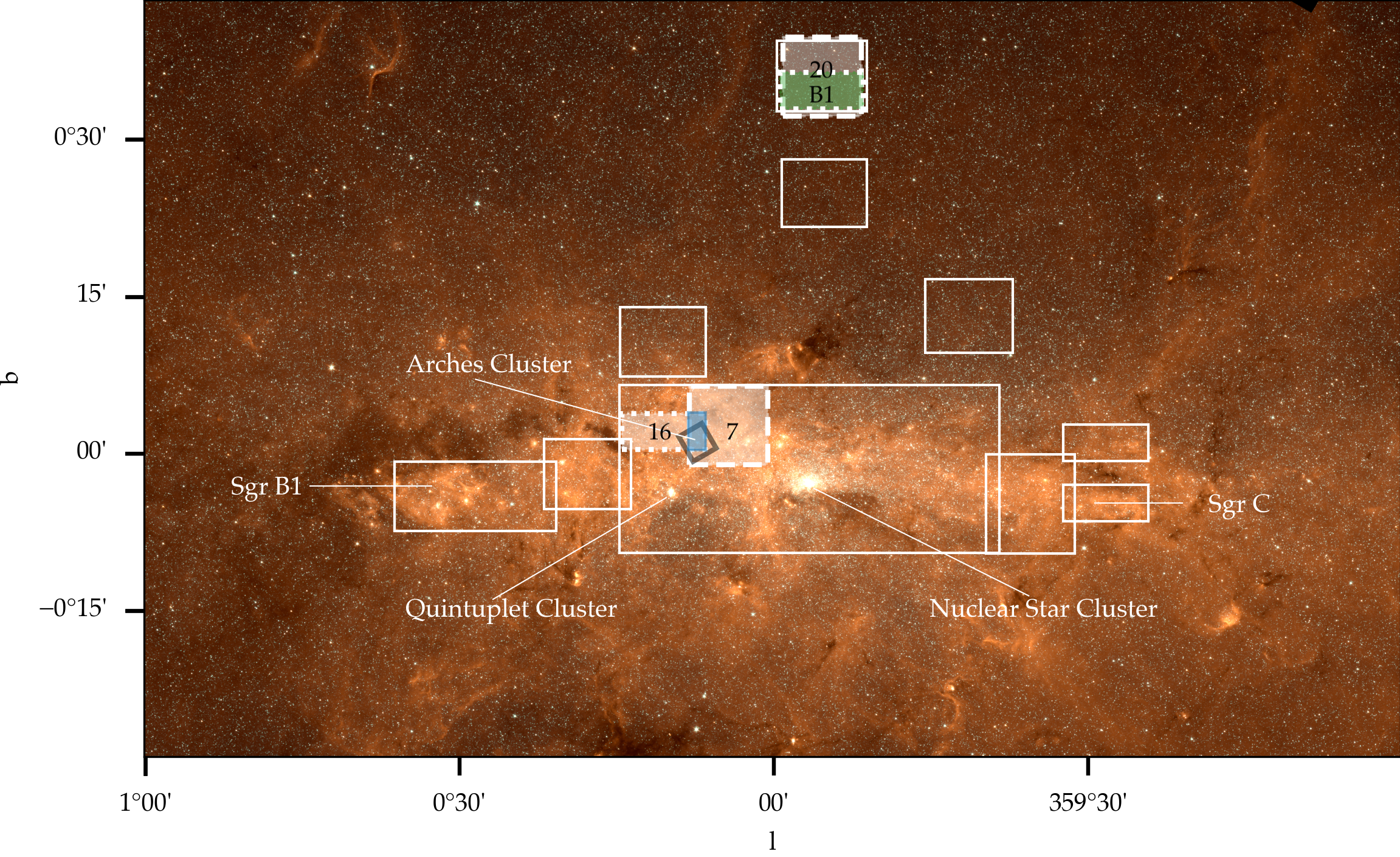}
		\caption{GNS survey fields overlaid on a Spitzer/IRAC colour mosaic \citep[3.6, 4.5, and 8\,$\mu$m;][]{Stolovy:2006fk}.White solid lines indicate the full extent of the GNS. White shaded regions mark the fields analysed in this work, while dotted and dashed outlines denote the fields of view of the GNS\,I and GNS\,II pointings, respectively. The black numbers label the individual fields (see Table~\ref{tab:fields}).Green and blue areas indicate the regions where we computed proper motions: the green area corresponds to a field on the Galactic bar, and the blue solid box corresponds to the GNS fields that overlap with the Arches cluster. 
        The black box shows the coverage of the catalogue from \cite{Hosek:2022om}.}
		\label{fig:gcview}
	\end{figure*}

In this paper, we present the new data reduction, analysis, and calibration pipeline for GNS\,1 DR2 and GNS\,II, and highlight the main improvements with respect to GNS\,I\,DR1 \citep{Nogueras-Lara:2019yj}. We also describe how we measured high-precision proper motions. To validate our methodologies, we analyzed two test fields with distinct characteristics, representative of the variety of environments across the GC. Specifically, we reduced observations of (i) an inner bar field located $\sim$0.6\,deg north of Sgr\,A* (field~B1  and field~20 in Fig.~\ref{fig:gcview}), and (ii) a highly crowded region within the nuclear stellar disk (field~16 and fiedl~7 in Fig.~\ref{fig:gcview}). The first field enables us to assess the overall performance of our pipeline in an environment that is less extinguished and crowded than the second field. In the latter case, we additionally examined the feasibility of stellar cluster detection by analyzing the area overlapping with the Arches cluster. The obtained properties for the Arches cluster compare very well with published values in the literature, thus further underlining the quality of GNS.


\section{Observations}

GNS\,I observations were obtained mostly in 2015, with additional observations in 2016 and 2017, with the wide-field near-infrared camera HAWK-I/VLT in fast-photometry mode. The data were reduced using the speckle holography algorithm \citep[][]{Schodel:2013fk}, achieving a homogeneous angular resolution of $0.2''$  \citep{Nogueras-Lara:2018pr}. The survey covers an area of $\sim6000$\,pc$^{2}$ (Fig.~\ref{fig:gcview}). Due to the extreme crowding in the GC, the sky background was estimated using dithered exposures of a dark cloud near the Galactic Center ($\alpha \approx 17^\mathrm{h}48^\mathrm{m}01.55^\mathrm{s}$, $\delta \approx -28^\circ59'20''$), where the stellar density is very low. For further details on these observations, we refer the reader to \cite{Nogueras-Lara:2019yj}.  GNS\,I\,DR1 provides  photometry for $\sim 3.3 \times 10^{6}$ stars in the $J$, $H$, and $K_s$ bands, with typical uncertainties of $\lesssim 0.05$\,mag in all three bands.

The observations for GNS\,II were acquired in 2022 with a  general observing strategy similar to the one used in GNS\,I.  
There are, however, two key differences between the two epochs. The first concerns the detector size: in GNS\,I, the fast-photometry mode was employed with a DIT of 1.26\,s, which restricted the usable area to one-third of the detector ($2048 \times 768$ pixels). In GNS\,II, with a DIT of 3.32\,s we were able to use the full detector array ($2048 \times 2048$ pixels), thus allowing us to image  three times larger areas with a single pointing. The second major difference is the use of the GRAAL ground layer adaptive optics system \citep[][]{Paufique:2010zl} in GNS\,II. 
We increased the DIT and used the full detector window to enable the use of the GRAAL adaptive optics system, thereby improving  Point Spread Function (PSF) stability across the field

In this pilot study, we have reduced and analyzed two fields for GNS\,I and two for GNS\,II. Details of these fields are shown in Table~\ref{tab:fields}. The dotted and dashed boxes in Fig.~\ref{fig:gcview} represent the fields from GNS\,I and GNS\,II, respectively.

\begin{table}[h!]
\centering
\small 
\caption{Observing detail of the fields reduced for this study.}
\begin{tabular}{lcccccc}
\hline\hline
Filter & Epoch & Field & Date & Seeing\tablefootmark{a} & DIT  \\
       &  &     & (d/m/y) & (arcsec) & (s) &  \\
\hline
$J$   & &    & 24/07/2015 & 0.43 &  \\
$H$   &1& B1 & 20/05/2016 & 0.56 & 1.26 \\
$K_s$ & &    & 28/06/2015 & 0.54 &  \\
$J$   & &    & 10/06/2015 & 0.48 & \\
$H$   &1& 16 & 10/06/2015 & 0.46 & 1.26 \\
$K_s$ & &    & 10/06/2015 & 0.55 &  \\
$H$   &2& 20 & 02/08/2022 & 0.52 & 3.32\\ 
$H$   &2& 7  & 27/05/2022 & 0.53 & 3.32  \\
\hline
\end{tabular}
\tablefoot{
\tablefoottext{a}{Average seeing during the observations.}
}
\label{tab:fields}
\end{table}

\section{Data reduction and analysis}

\subsection{Data reduction pipeline}

In this section we describe our data reduction procedures up to the point of speckle holography. We highlight similarities and differences with GNS\,I\,DR1. We processed the data from  each of HAWK-I's four detector chips  separately.

\begin{figure}
    \centering
    \includegraphics[width=1\linewidth]{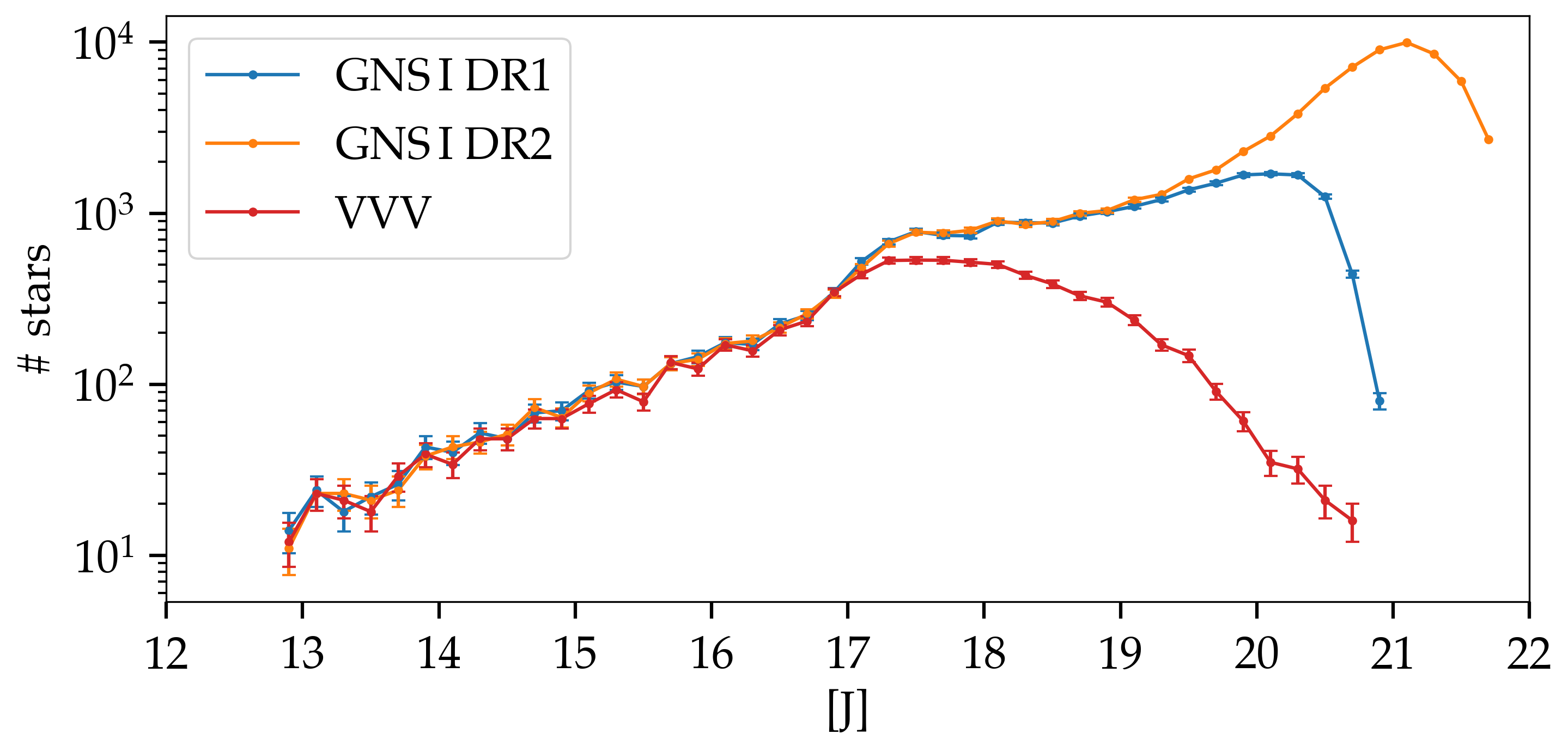}
    \includegraphics[width=1\linewidth]{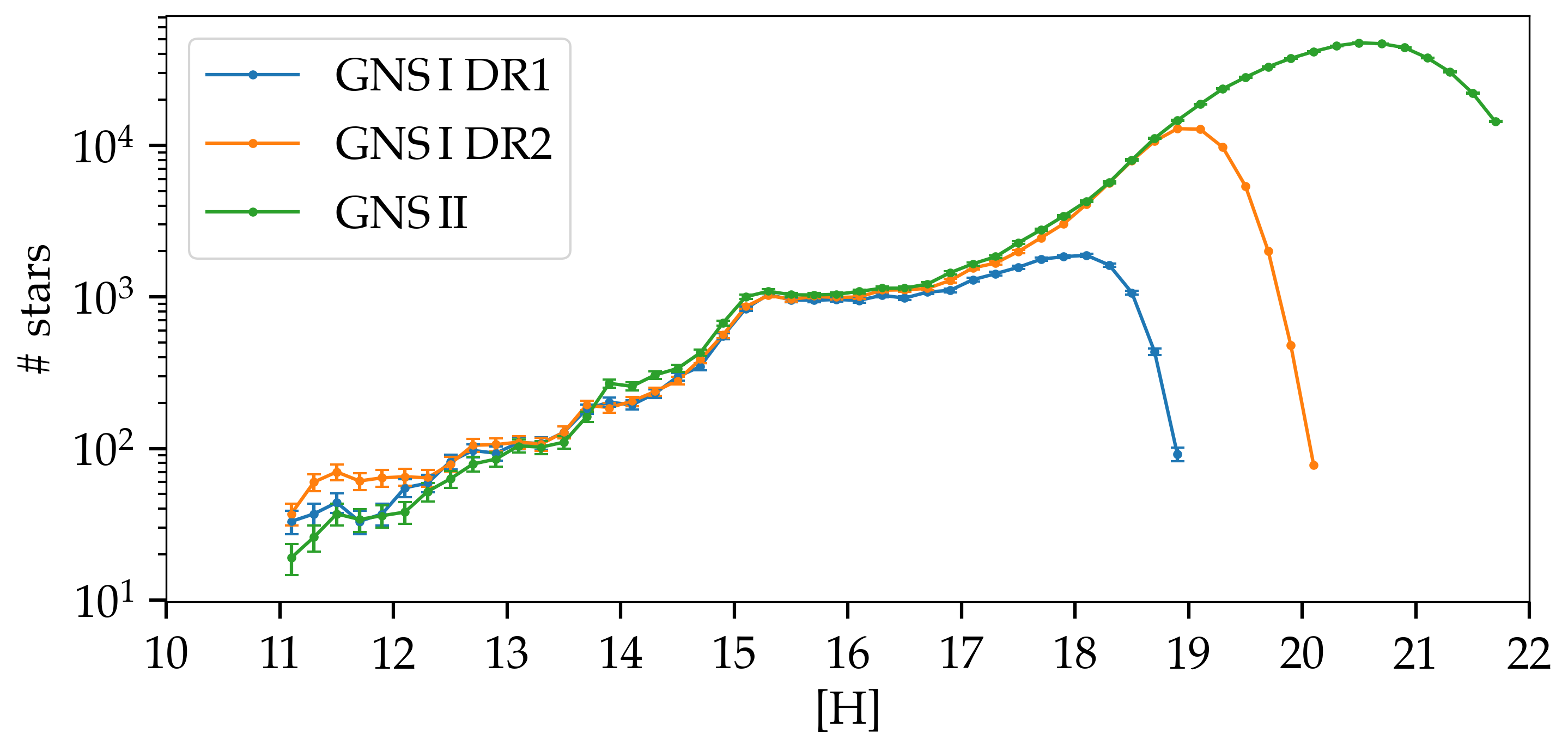}
    \includegraphics[width=1\linewidth]{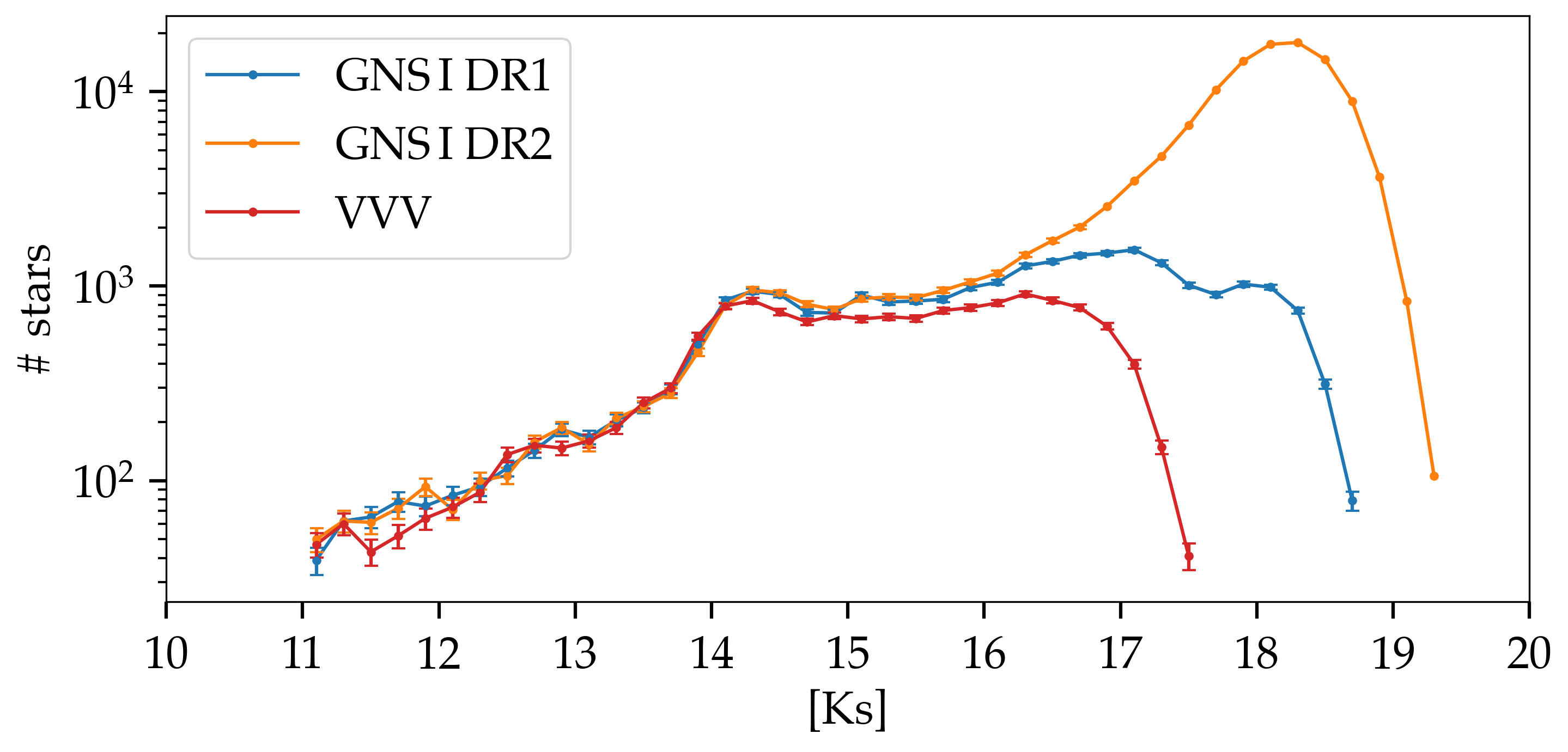}
    \caption{Comparison of the $J$, $H$, and $K_s$ luminosity functions for the Galactic bar field (green box in Fig.~\ref{fig:gcview}), using data from the VVV catalog by \cite{Griggio:2024eb} (available only for the $J$ and $K_s$ bands), as well as from GNS\,I (DR1 and DR2) and GNS\,II.}
    \label{fig:l-functions}
\end{figure}

\begin{enumerate}
    \item Bad pixel correction, flat-fielding, and sky subtraction were carried out as in GNS\,I\,DR1 \citep{Nogueras-Lara:2019yj}. Contrary to GNS\,I\,DR1, we  used the median of the pixels in the lowest 5\%  range of values of each dark subtracted science frame to scale the normalised sky before its subtraction. It was 10\% in case of DR1. This largely avoids negative pixels in the reduced science frames.
    \item Deselection of bad frames. Sometimes the telescope moved during the exposures, resulting in images with smeared or duplicated stars. We rejected those images.  On the AO data (GNS II) We used the MaxiTrack tool \citep{Paillassa:2020ys}, a convolutional neural network designed to automatically identify tracking and guiding errors in astronomical images. This tool does not work well on the older speckle data (GNSI), for which we identified the bad frames by eye.
    \item Geometric distortion correction and precise relative alignment of all short exposures, as described in detail in the next section. Geometric distortion correction was done with respect to a VVV image in GNS\,I\,DR1. For DR2 we used the SCAMP and SWarp software packages from the \href{https://www.astromatic.net/}{Astromatic}  site \citep{Bertin:2002la,Bertin:2006fp}. The SExtractor package was used to support these programmes \citep{Bertin:1996oq}. This procedure is a significant change compared to DR1 and this step has proven to be essential to reach the high astrometric accuracy of GNS\,I\,DR2 and GNS\,II. 
    \item Creation of a long exposure image with its corresponding noise image from the mean and error of the mean of the individual short exposures. The StarFinder package \citep{Diolaiti:2000fk} was used to extract stars and their photo-astrometry from the long-exposure to use them in the holographic image reconstruction \citep[see below and][]{Schodel:2013fk,Nogueras-Lara:2018pr}. 
    \item Creation of  image cubes for $1\times1$\,arcmin$^{2}$ sub-regions as in GNS\,I\,DR1. The sub-regions overlap by $0.5$\,\arcmin on each side, except at the edges of the field-of-view. The cubes contain all individual short exposures (of length DIT) corresponding to this region.
    \item Speckle holography reduction of the sub-regions. Different than in GNS\,I\,DR1, where three sub-images were created from disjunct data, we created a deep image with all frames plus ten jack-knife sampled images, which left each out a different 10\% of the frames. The advantage here is that we can obtain deeper  photometry (see Fig. \ref{fig:l-functions}) while maintaining reliable noise estimates. As in DR1 we resampled all exposures by a factor of two, using cubic interpolation. A Gaussian beam of $0.2"$ FWHM was used for beam restoration in the speckle holograpjhy algorithm. This resulted in final images with an excellent sharpness and a homogenous Gaussian PSFs. In the following we refer to the reconstructed $1'\times1'$ images as sub-images.
\end{enumerate}

\subsection{Geometric distortion correction}

	We used \href{https://www.astromatic.net/software/scamp/}{SCAMP} \citep{Bertin:2006fp} to correct for geometric distortion and to compute the global astrometric solution. We only provide a brief description of the algorithm here. The interested reader can find more details about how SCAMP works in \citet{Bouy:2013lg}.
 
    SCAMP is fed with position lists extracted from each exposure with \href{https://www.astromatic.net/software/sextractor/}{SExtractor}\citep{Bertin:1996oq}. It computes the global geometric solution by minimizing the squared positional differences between overlapping sources ($\chi^2_{\mathrm{astrom}}$) in pairs of catalogs:
	
	\begin{equation}
		\chi^2_{\mathrm{astrom}} = \sum_s \sum_a \sum_{b > a}
		\frac{ \left\| \boldsymbol{\xi}_a(\mathbf{x}_{s,a}) - \boldsymbol{\xi}_b(\mathbf{x}_{s,b}) \right\|^2 }
		{ \sigma_{s,a}^2 + \sigma_{s,b}^2 }
		\label{eq:1}
	\end{equation}

where, $s$ indexes the matched sources, while $a$ and $b$ denote different images. The quantity $\mathbf{x}_{s,a}$ represents the observed position of source $s$ in catalog $a$, typically in pixel coordinates. The function $\boldsymbol{\xi}_a(\mathbf{x}_{s,a})$ is the transformation that maps these coordinates into a common astrometric reference frame using the current calibration parameters for catalog $a$. The terms $\sigma_{s,a}$ and $\sigma_{s,b}$ denote the positional uncertainties associated with source $s$ in catalogues $a$ and $b$. 
	
	\begin{figure}
		\centering
		\includegraphics[width=1\linewidth]{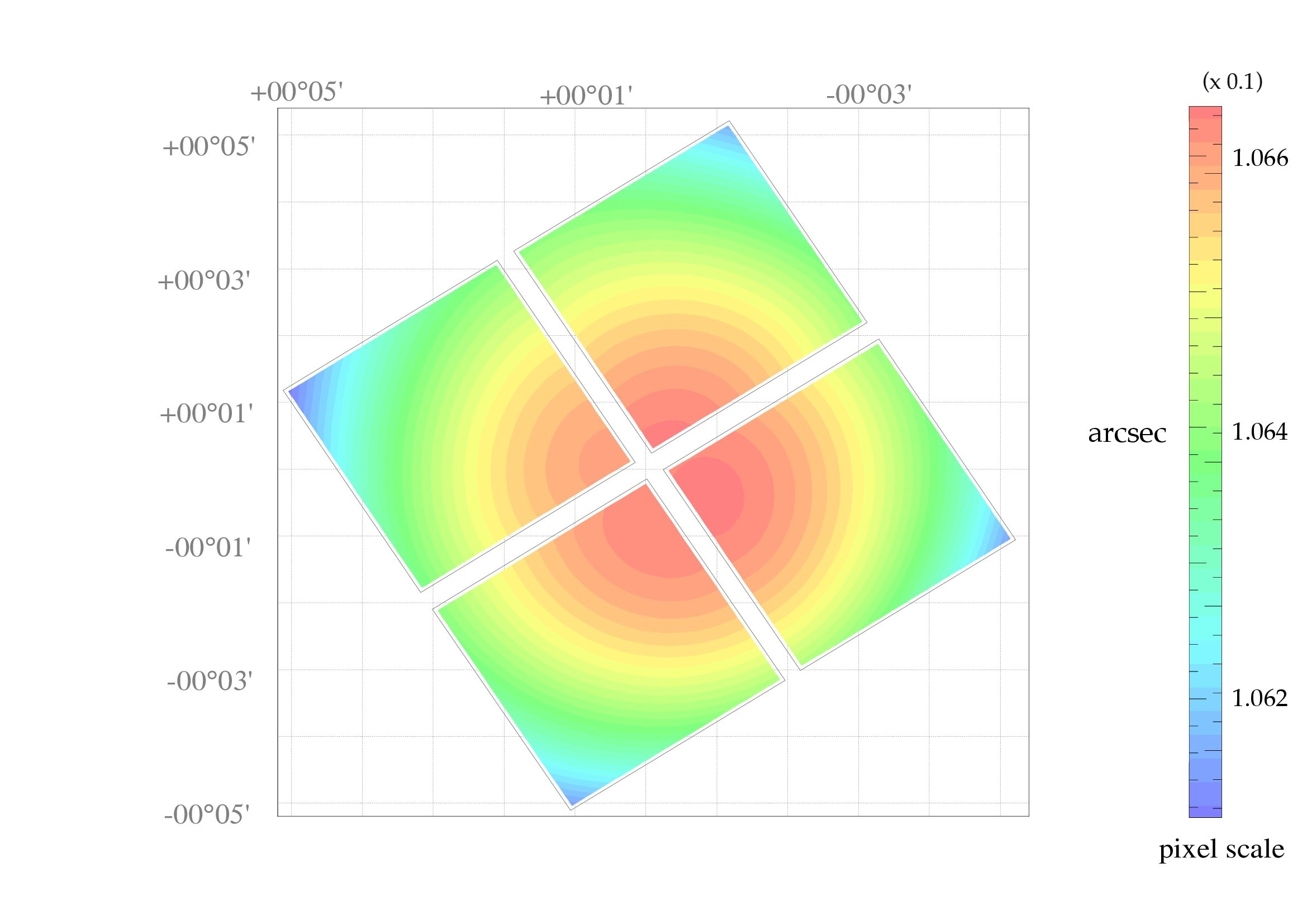}
		\caption{Example of the HAWK-I mosaic camera distortion map provided by SCAMP for a GNS\,II pointing. Colors encode the variation of the pixel scale across the detector.}
		\label{fig:scamp_exm}
	\end{figure}

    We show in Fig.~\ref{fig:scamp_exm} an example of the distortion pattern determined by SCAMP for the HAWK-I camera from a GNS\,II pointing. The color variations in the image represent the deviations of the local pixel scale from the nominal one. 
    
    Based on the position catalogs derived from each image, SCAMP generates updated headers for each exposure that encode the geometric distortion solution. These corrected headers are then applied to the images, which are subsequently re-projected onto a common grid using \href{https://www.astromatic.net/software/swarp/}{SWarp} \citep{Bertin:2002la}. After processing with SWarp, all exposures share a consistent astrometric solution and are resampled onto a common, distortion-corrected reference frame.
    
\subsection{Photometry and astrometry}\label{Photometry and astrometry}

Subsequently, we performed point source fitting on the holographically reconstructed images of the $1\times1$\,arcmin$^{2}$ sub-regions and created lists and images for the full field-of-view of each pointing.

\begin{figure*}
    \centering
    \includegraphics[width=\textwidth]{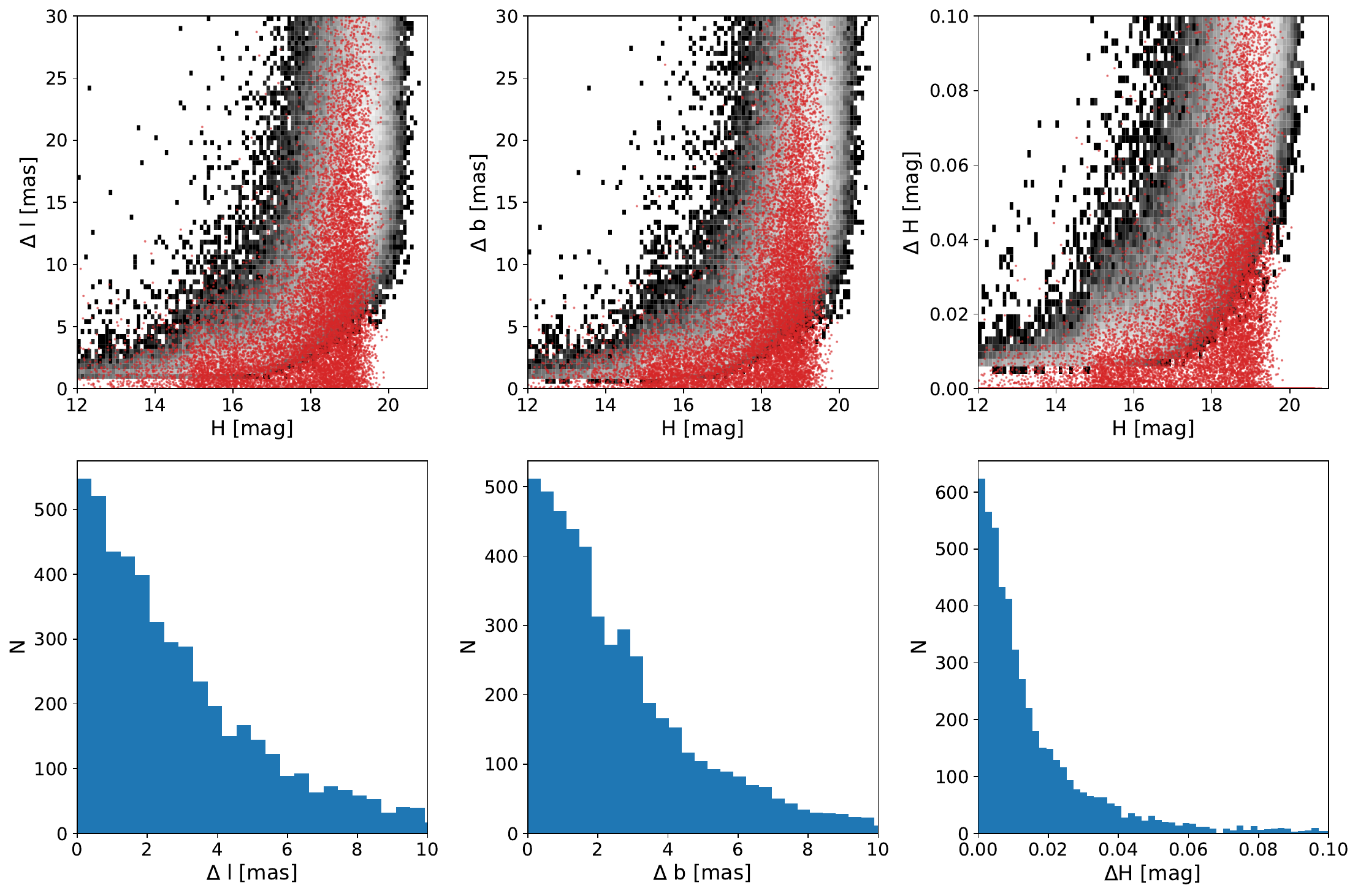}
    \caption{Astrometric and photometric uncertainties of the stars in the Galactic bar field (green box in Fig. \ref{fig:gcview} using dara from GNS\,I\,DR2. Upper row: The gray shaded 2D histograms show the uncertainties estimated by the pipeline for stars of a given magnitude (combined jackknife and PSF uncertainties). The red dots show the uncertainties of stars as estimated from their multiple detections on different detector chips. Lower row: Histograms of the astrometric and photometric uncertainties of bright ($H\leq18$~[mag]) stars measured on different detectors.}
    \label{fig:uncertainties}
\end{figure*}

\begin{enumerate}
    \item The PSF of each sub-image is extracted with an automatic script, based on StarFinder, that builds a median PSF from five to 40 unsaturated, bright, and isolated stars (any star within $0.4"$, about the FWHM of the PSF of a reference star must be at least five magnitudes fainter; the exact numbers can be varied without any significant impact on the results). The PSF is extracted iteratively to take into account secondary sources near the reference stars. Five stars is the minimum necessary for a good PSF determination (and to minimize bias due to individual stars). While about 20 reference stars will provide a very good estimate of the PSF, up to 40 are used in case a field is extremely crowded or contains only faint stars. The exact number can be varied and does not have any significant influence on the results.
    \item The positions and fluxes of the stars are measured with StarFinder with two iterations, using  relative thresholds of $3\,\sigma$, and a minimum correlation  of $0.7"$. Extended emission may be present in the images due to unresolved stellar sources or diffuse emission from ionised gas. It is subtracted by StarFinder before making the measurements. The extended emission is estimated by StarFinder on $2.4"\times2.4"$ regions and then interpolated across the field. We have chosen this region large enough to obtain reasonably accurate estimates of the diffuse emission in the crowded GC. It does not have any strong influence on the results because StarFinder will always fit a slanted background when fitting the core of any star with the PSF.
    \item The measurement process is carried out on the deep image and its corresponding jackknife sampled images. The resulting point source lists are then combined. A source is considered real only if it is detected in the deep image and in all jackknife images. The uncertainties are determined according to standard jackknife statistics: For $n_{jack}$ samples, $x_{i}$, the uncertainty of the mean, $\overline{x}$, is $\sqrt{\Sigma{(x_{i}-\overline{x})^{2}} \times (n_{jack}-1)/n_{jack}}$.
    \item Median astrometric and photometric offsets between the source lists of pairwise overlapping sub-images are determined from all sources brighter than 17\,mag (all bands) that coincide within $0.1"$, and that are located at least $1"$ from the sub-image edges (to avoid potential edge effects), typically between  100-200 sources. These offsets can be related to small variations of the estimated PSFs between the sub-images.
    \item The best astrometric and photometric offsets for each sub-image and their corresponding source
    lists (sub-lists) are estimated with a global optimization algorithm that minimizes the mean square deviations between the pairwise overlapping sub-regions. This procedure is basically the one that is described in Appendix~A of \citet{Dong:2011ff} and is a major difference to GNS\,I\,DR1, where all measurements were related to a single sub-region and not globally optimized. This global optimization is another key tool in achieving the high photometric and astrometric accuracy of GNS\,I DR2 and GNS\,II.
    \item The astrometric offsets determined in the previous step are applied to the sub-images and their sub-lists to reconstruct a large image and source list for each chip. At this point the pipeline provides two kinds of uncertainties, that is (1)  statistical uncertainties from the jackknife sampling and (2)  uncertainties due to variations between the sub-images that can be estimated from the two to four measurements for sources in the overlap regions, caused by, e.g.\ uncertainties in the PSF extraction. This uncertainty is assumed to be the same for all stars. We determine the median of the latter uncertainties for bright ($J<17, $$H<16$, $K_{s}<16$), unsaturated stars and add it in quadrature to the jackknife uncertainties to take into account this systematics for all sources. Thus we introduce a floor uncertainty for all sources, which ranges from $0.5-1.5$\,mas in position and $0.02-0.05$\,mag in photometry.
    \item Astrometric calibration was done with recently published high precision measurements of the VVV survey towards the GC \citep{Griggio:2024eb}.
    \item Photometric calibration was based on the SIRIUS/IRSF survey \citep{Nagayama:2003fk,Nishiyama:2006ai}, as in GNS\,I\,DR1. We chose the reference stars to be bright, unsaturated and isolated (any other star within a $2"$ radius must be at least 5\,mag fainter). We obtained about 100 (GNS\,I) to 300 (GNS\,II) stars per chip for photometric calibration, which established the zero point with respect to the SIRIUS/IRS survey with (sub)percent precision. 
    \item Finally, any astrometric and photometric offsets between the four detector chips were corrected by computing pairwise mean offsets and finding the optimal offset for each chip with a minimization routine, in the same way as we did for the sub-regions. To avoid any bias introduced by this step, we re-calibrated the photometry then again  with SIRIUS (see step before). The astrometry of the final lists was calibrated with respect to Gaia DR3 sources within the field, applying simple shifts in Galactic latitude and longitude.
\end{enumerate}

\section{Astrometric and photometric performance}

The pipeline determines relative astrometric and photometric uncertainties in two ways, that is (1) from the uncertainties estimated by jackknife sampling and (2) from the uncertainties estimated from multiple measurements of stars in overlapping parts of the sub-images. We refer to the  latter uncertainties as PSF uncertainties, because they are probably mostly limited by the precision with which the PSF can be estimated for each sub-image, but other sources of uncertainty, such as uncertainties of the flat field, will play a role, too. The PSF uncertainties provide a lower floor to all measurements. We therefore add them in quadrature to the jackknife uncertainties.

Due to the dithering of the observations, a large number of stars are observed multiple times on different detector chips. We can use these independent measurements to cross-check the robustness of the uncertainties estimated by our pipeline. Fig.~\ref{fig:uncertainties} shows a summary plot. The upper panel compares the uncertainties estimated by the pipeline (2D histograms) to the uncertainties from multiple measurements on different detectors (red dots). The comparison shows that our pipeline provides robust, possibly even slightly overestimated uncertainties. The uncertainties estimated from the independent measurements on different chips may be underestimated at times, because there are only two measurements for many stars. We conclude that the uncertainties provided by the pipeline are realistic and provide reliable minimum uncertainties at each magnitude. The lower plot shows histograms of the uncertainties of bright ($H\leq18$\,mag) stars detected on different detector chips. We omit faint stars to minimize the influence of random uncertainties. The histograms show astrometric uncertainties of a few milliarcseconds and photometric uncertainties  of a few percent. These results are consistent across the $J$ and $K_s$ bands, as well as for the GNS\,II data.

\begin{figure}
    \centering
    \includegraphics[width=1\linewidth]{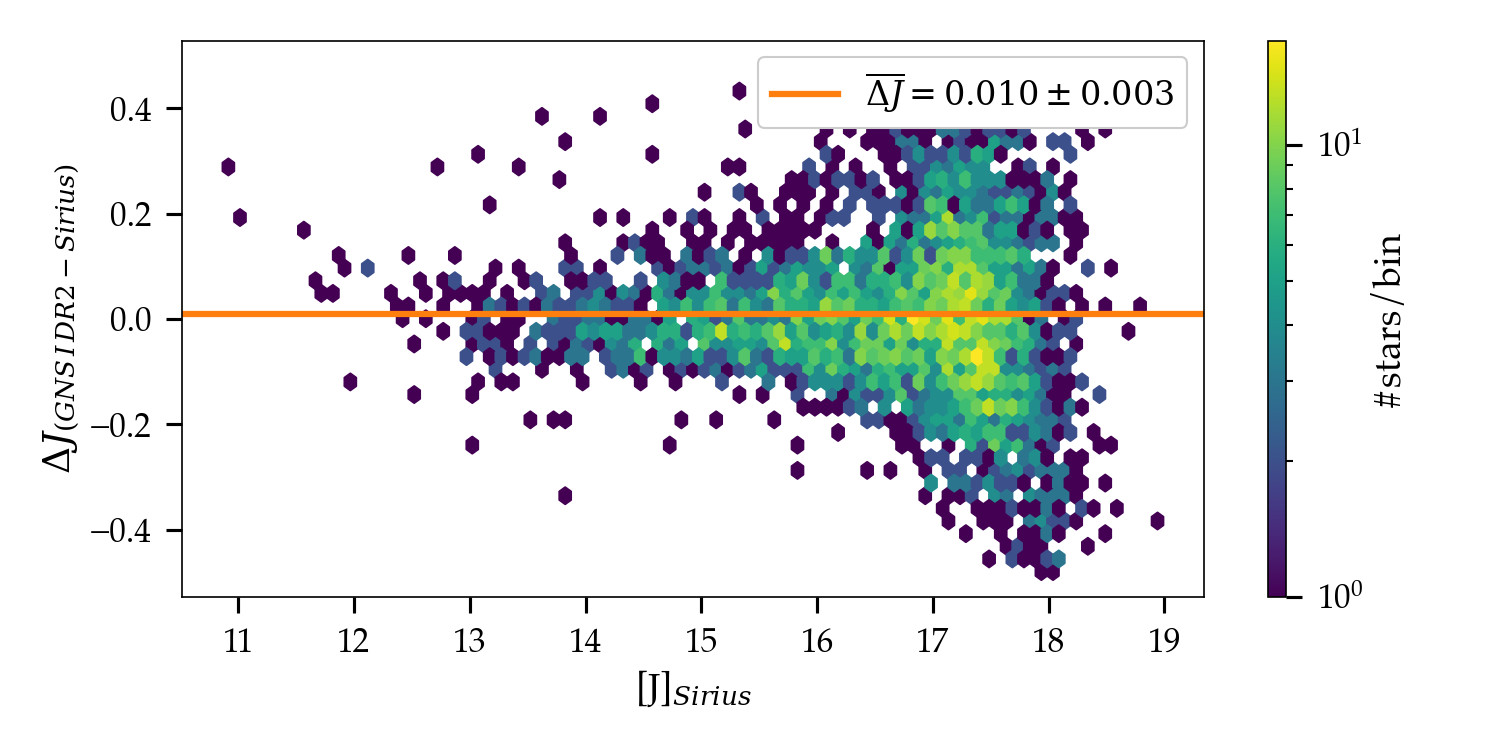}
    \includegraphics[width=1\linewidth]{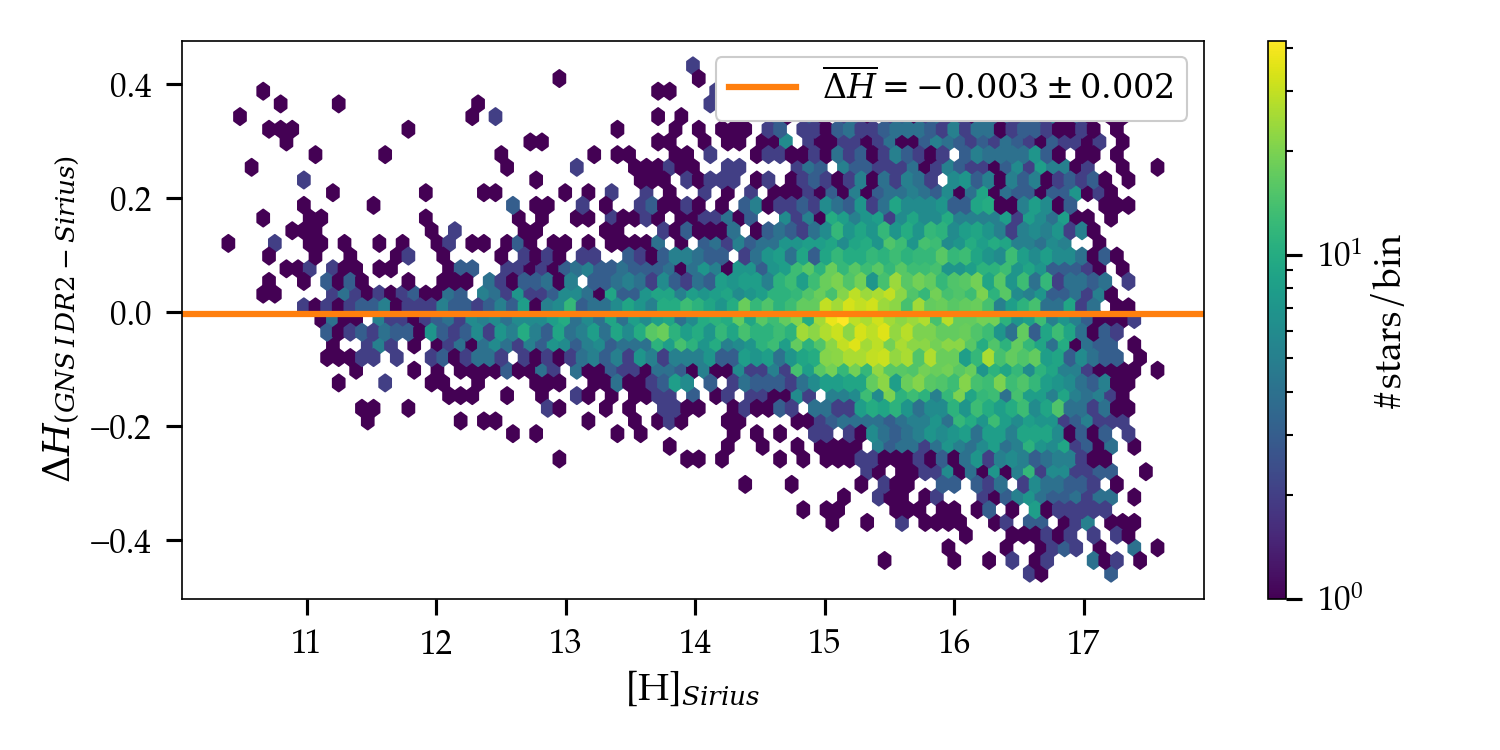}
    \includegraphics[width=1\linewidth]{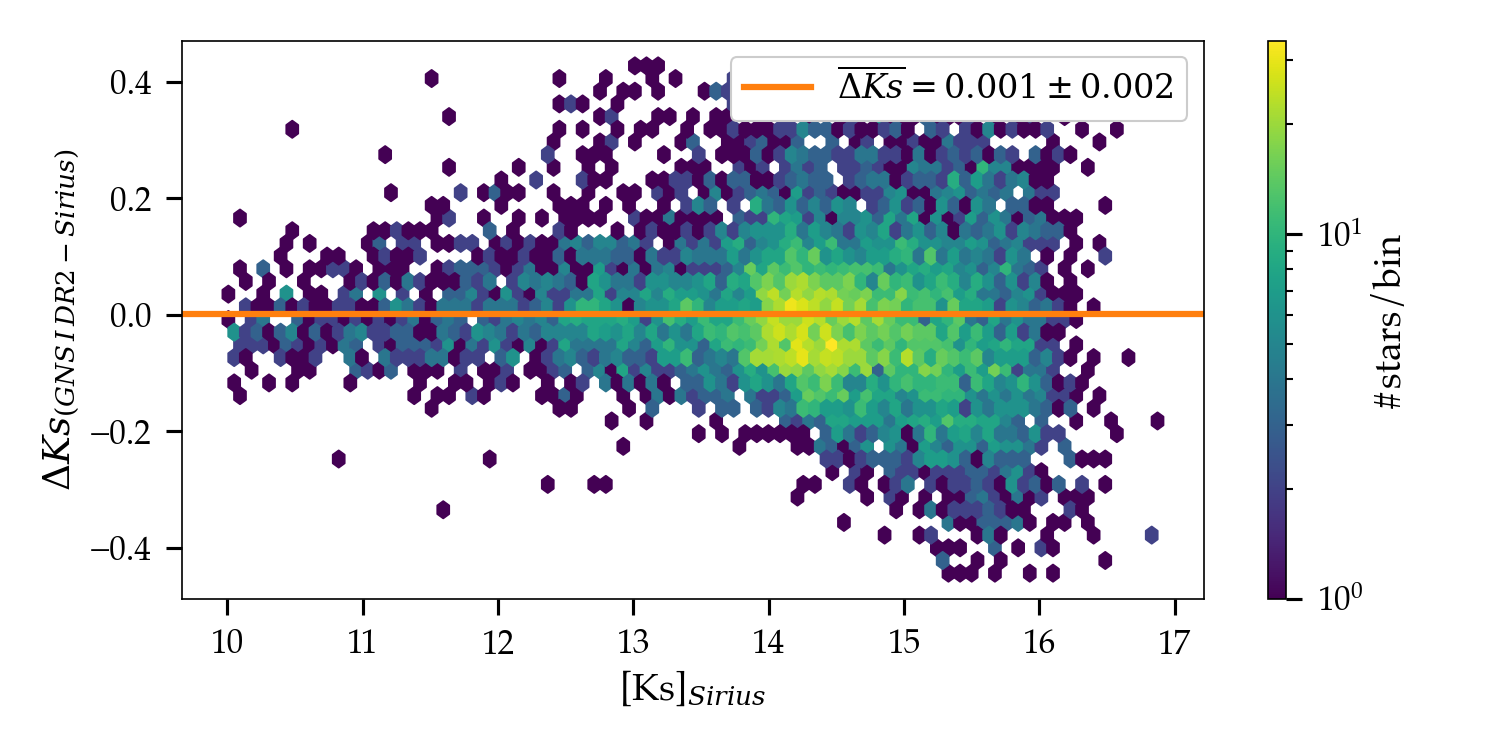}
    \caption{Magnitude residuals between GNS\,I\,DR2 and SIRIUS for the $J$, $H$, and $K_s$ bands in the Galactic bar field (see Fig.~\ref{fig:gcview}). The orange line indicates the mean of the residuals.}
    \label{fig:photometry}
\end{figure}

Figure\,\ref{fig:photometry} compares the photometry of GNS\,I\,DR2 in the $J$, $H$, and $K_s$ bands with that of the SIRIUS survey for all common stars in the field~B1. The mean deviations and their uncertainties are $0.010\pm0.002$\,mag for the $J$ band, $0.003\pm0.002$\,mag for the $H$ band, and $0.001\pm0.002$\,mag for the $K_s$ band. This indicates that the uncertainty of the GNS\,I\,DR2 photometry is dominated by the 3\% systematic zero-point uncertainty of the SIRIUS survey \citep{Nishiyama:2005ao}. As shown in Fig.\,\ref{fig:photometry} middle panel, saturation begins to affect the photometric measurements in GNS\,I at magnitudes $H\lesssim11$ \citep[see also][]{Nogueras-Lara:2019yj}. For GNS\,II, the residuals relative to SIRIUS are fully consistent, with a mean deviation of $0.001\pm0.002$\,mag in the $H$ band.

\begin{figure}
    \centering
    \includegraphics[width=1\linewidth]{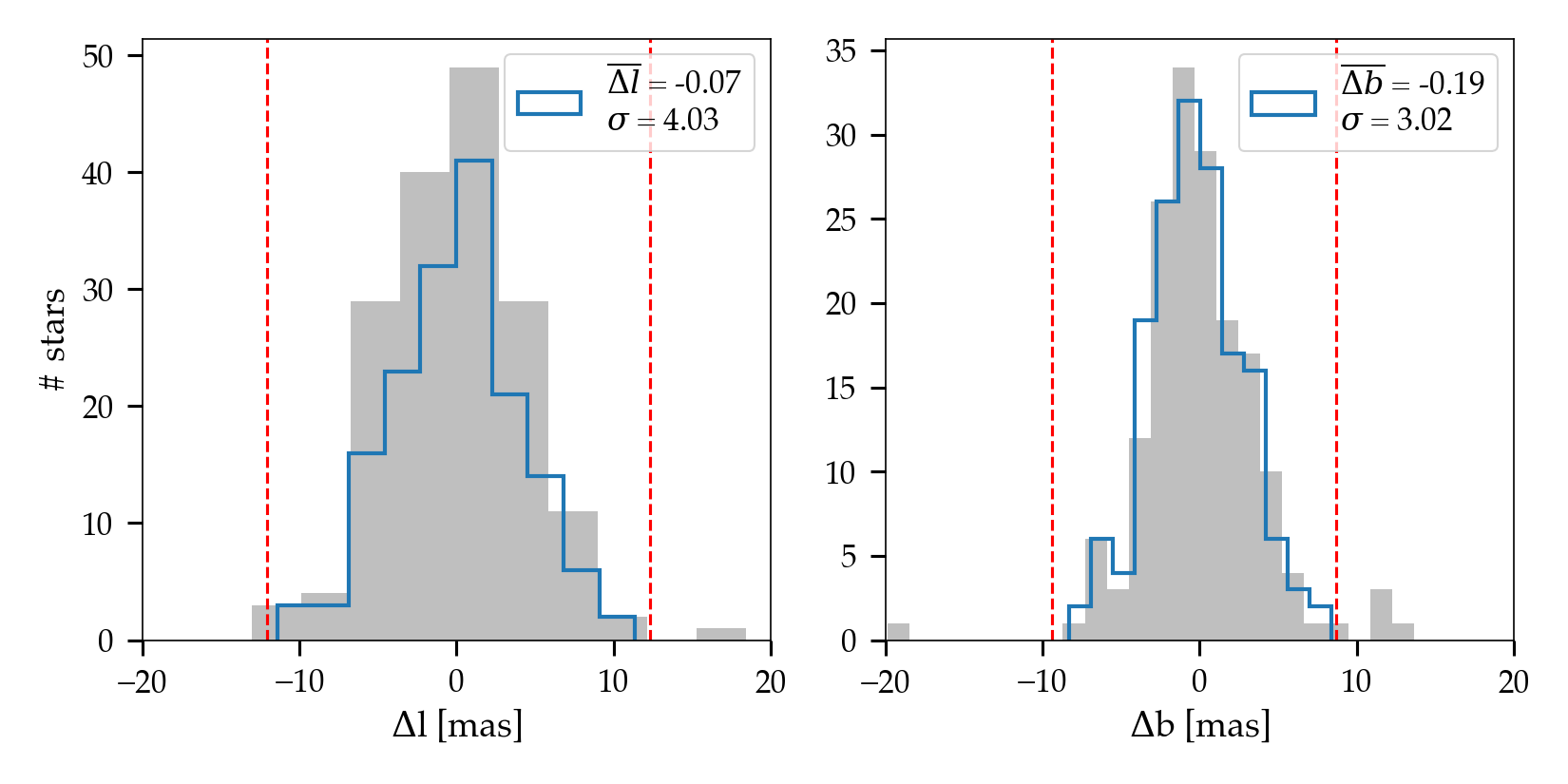}
    \caption{Position residuals in $l$ and $b$ between GNS\,I\,DR2 $H$ band stars in the field~B1 ( Fig.~\ref{fig:gcview}) and Gaia DR3 stars in the same field. Blue histograms show the residual distribution after removing the $3\sigma$ outliers, while gray histograms show the full distribution before clipping. Red dotted lines indicate the $3\sigma$ thresholds.}
    \label{fig:astrometry}
\end{figure}

To estimate the astrometric accuracy, we calculated the position residuals in $l$ and $b$ for the $H$ band with respect to Gaia~DR3 \citep{Gaia-Collaboration:2023lr} for field~B1. Common stars were selected by searching for cross-matches within $0.05''$. Subsequently, a similarity transform was applied to remove a small systematic offset ($16.5$\,mas in $l$ and $3.3$\,mas in $b$). As shown in Fig.\,\ref{fig:astrometry}, the residual differences between GNS\,I\,DR2 and Gaia\,DR3 exhibit a standard deviation $<4$\,mas in both $l$ and $b$. Similar residuals with Gaia~DR3 are found when cross-matching with GNS\,II. Performing the same residual comparison with Gaia~DR3 for the same field in GNS\,I\,DR1 yields values of about $20$\,mas, indicating an improvement by roughly a factor of five with respect to GNS\,I\,DR1.

\section{Proper motions}


In order to compute proper motions, the stellar positions need to be referenced to a common coordinate system. Although our absolute astrometric uncertainties are already very small, achieving precise proper motions requires defining a stable reference frame that minimizes the transformation uncertainties between epochs. 
To this end, we employed two different and independent methodologies to define the reference frame and compute the proper motions, which we refer to as relative proper motions and absolute proper motions. For the proper motion calculations, we used stars in the $H$ band from each epoch.


\subsection{Relative proper motions}\label{sec:relative_pm}

In the relative method, no absolute reference frame (such as the International Celestial Reference System, ICRS) is used. Instead, one epoch of the GNS dataset is adopted as the reference epoch, and the stellar positions from the other epoch are aligned to this frame. If a sufficiently large number of stars is available, one can assume that their individual proper motions cancel out statistically, so that the mean stellar motion is effectively zero. This allows the construction of a stable reference frame against which the relative proper motions of individual stars are measured. This technique has been applied in a large number of studies of the GC \citep[][]{Eckart:1997jl,Ghez:1998ad,Schodel:2009zr,Shahzamanian:2022vz,Martinez-Arranz:2022wj}.
    
Here we use GNS\,II as the reference epoch. The results obtained when using GNS\,I\,DR2 as  reference epoch are fully compatible. The final product of the data reduction and analysis pipeline is a list of stars for each pointing, obtained by combining the four detector chips. Considering the nominal size of the HAWK-I detector and the jittering applied during the observations, each of these lists covers an area of $\sim 7.8'\times 3.5'$ for GNS\,I\,DR2 and $\sim 7.8'\times 7.8'$ for GNS\,II.

We projected the stellar coordinates from both epochs onto the same tangential plane. The GNS\,I\,DR2 sources were then aligned with those of GNS\,II. To select suitable reference stars and ensure uniform sampling, we divided the reference field into a grid of $\sim 2.5 \times 2.5$ arcsec, thereby avoiding a bias toward regions of higher stellar density. In each cell, we selected a single reference star with $12 < H < 18$\,mag, choosing the one with the lowest positional uncertainty and ensuring it was isolated within a radius of $1''$ from any companion.

We cross-matched the reference set with the target epoch, considering a positive match as a source within 50\,mas and with an $H$-band magnitude difference within $3\sigma$. A similarity transformation was then computed from the matched sources and applied to the entire target catalog. The catalogs were cross-matched again, and a first-degree polynomial transformation was determined from the new matches and applied to all sources. This process was repeated iteratively until the number of matches stopped increasing.

Subsequently, the polynomial degree was increased to two, and the iterative alignment procedure was repeated.  Little to no improvement was achieved with a third-degree polynomial, so we limited our procedure to a polynomial degree  of two. Finally, the aligned positions were compared with those in the reference frame, and the positional offsets were divided by the time baseline to derive the proper motions, with standard error propagation applied to estimate the uncertainties.

\begin{figure}[h!]
    \centering
    \includegraphics[width=1\linewidth]{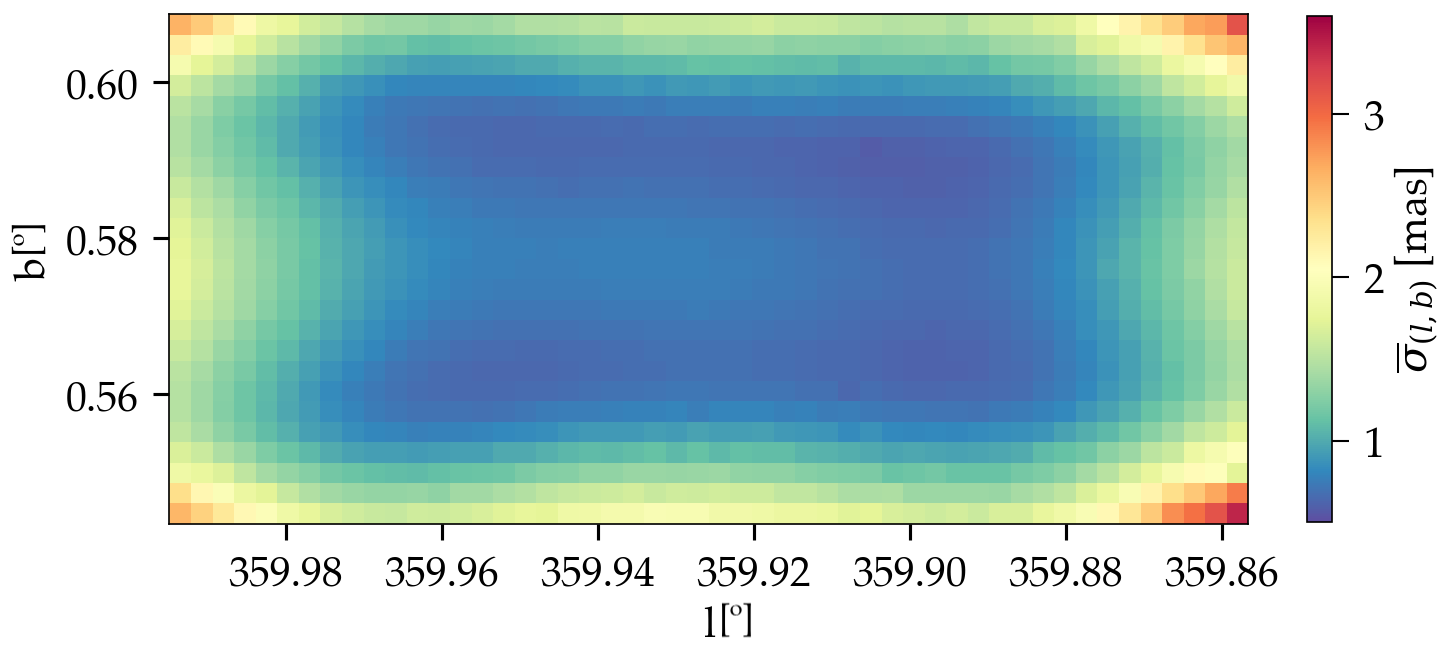}\\[1ex] 
    \includegraphics[width=1\linewidth]{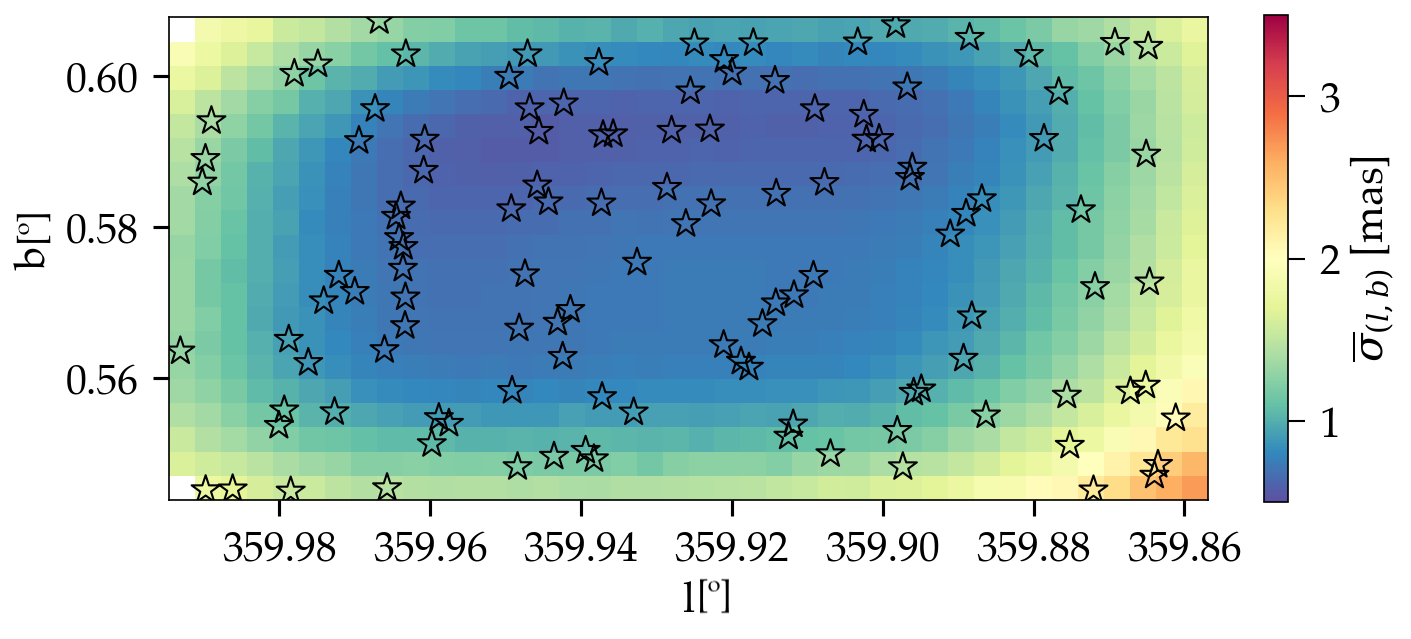}       
    \caption{Top: Total relative alignment uncertainties across the green region in Fig. \ref{fig:gcview}. Bottom: Mean absolute alignment uncertainties in the same region, with 
    Gaia reference stars marked in black.}
    \label{fig:B1_align_uncer}
\end{figure}

To estimate the uncertainty of the alignment procedure, we applied a bootstrapping method by randomly resampling (with replacement) the set of  reference stars 300 times. The standard deviation of the resulting distributions of the reference star positions was adopted as the estimate of the alignment uncertainty. The top panel of Fig.~\ref{fig:B1_align_uncer} shows the map of the  mean alignment uncertainty as a function of   $l$ and $b$ ($\overline{\sigma}_{l,b} = (\sigma _{l_{align}} + \sigma_{b_{align} })/2$). The alignment uncertainty is well below 2\,mas across almost the entire field. 


\begin{figure}
        \centering
        \includegraphics[width=1\linewidth]{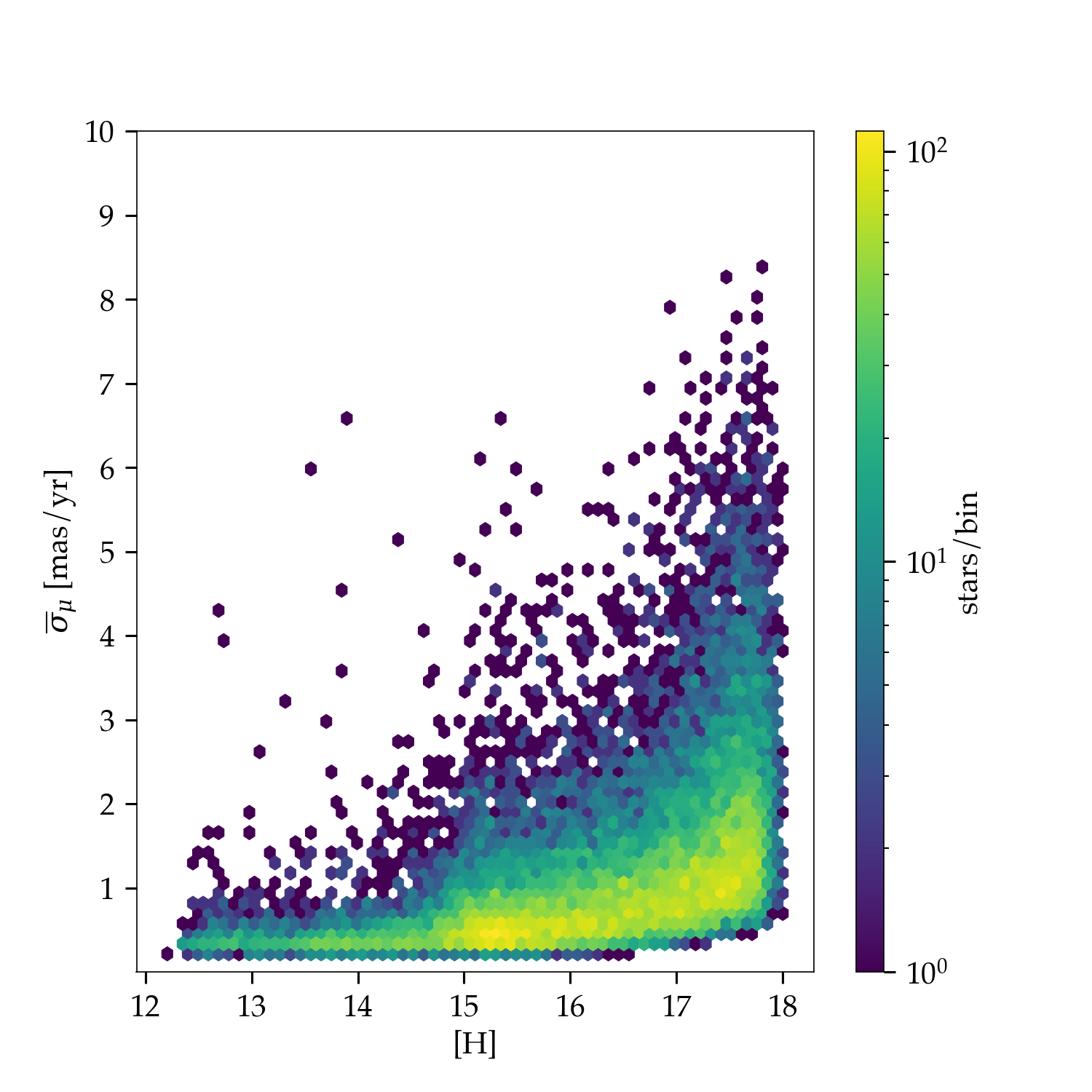}
        \caption{Mean proper motion uncertainty versus H magnitude from the relative alignment. This plot correspond to the stars in the green box region in Fig. \ref{fig:gcview} for stars with magnitues 12 < $H$ < 18.}
        \label{fig:pmE_vs_H}
    \end{figure}

    To determine the overall proper motion  uncertainties we applied standard error propagation by taking into account the individual position uncertainties for each star in the directions parallel and perpendicular to the Galactic plane for each epoch combined with the  alignment uncertainty. In Fig. \ref{fig:pmE_vs_H} we show the mean  proper motion uncertainty ($\overline{\sigma}_{\mu} = (\sigma_{\mu_{l}} + \sigma_{\mu_{b} })/2$) for the parallel and perpendicular components versus the H magnitude.

	Finally, to asses the quality of the relative proper motions, we compared our proper motions with those measured by Gaia DR3. We identified Gaia stars within our Galactic bar field (green box in Fig.\ref{fig:gcview}) and used their proper motions to calculate the expected positions of Gaia stars at the time of our reference epoch. We propagated the uncertainties accordingly. We transformed the proper motions to Galactic coordinates $\mu_{\ell} \times \cos b$ and $\mu_{b}$. We applied a quality cut to select the most suitable stars for the comparison: i) we excluded Gaia stars with magnitudes fainter than $G = 19$ and brighter than $13$, to avoid high astrometric uncertainties; ii) we discarded Gaia stars with a close Gaia companion to prevent mismatching; iii) we selected only sources with a 5-parameter astrometric solution (position, parallax, and proper motion); and iv) we eliminated Gaia sources with negative parallaxes, which is unphysical. 
    
\begin{figure}[h!]
    \centering
    \includegraphics[width=1\linewidth]{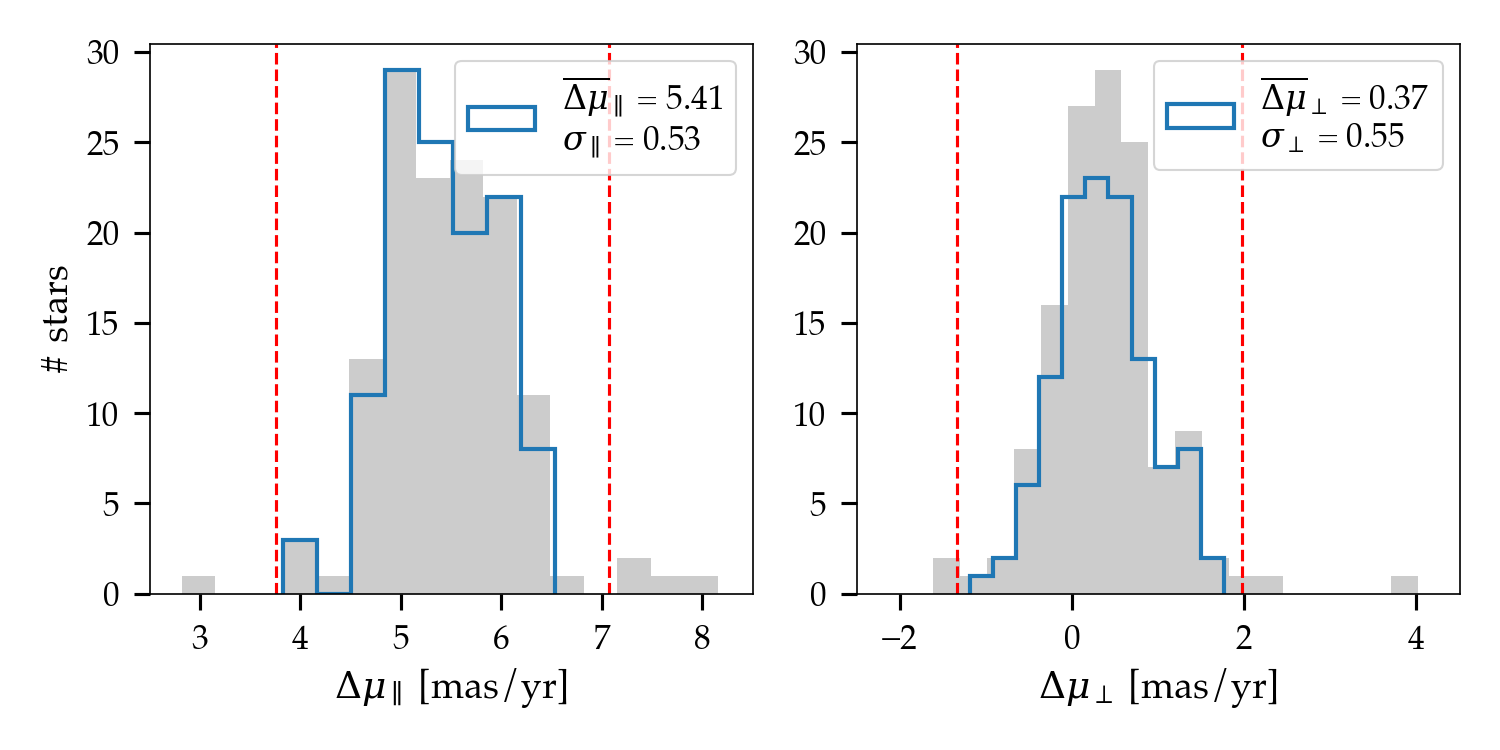}\\[1ex] 
    \includegraphics[width=1\linewidth]{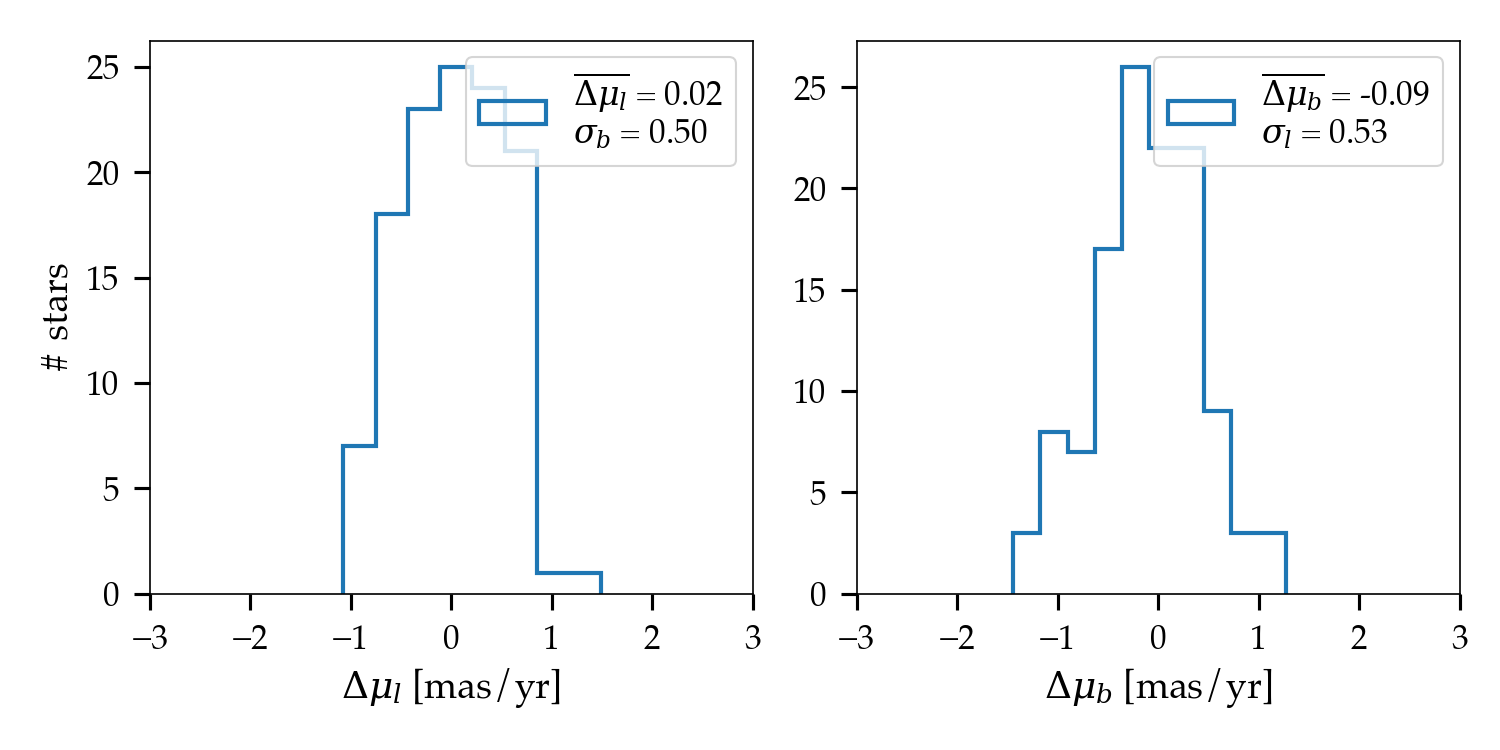}       
   \caption{Gaia--GNS proper motion residuals for the Galactic bar field (green box in Fig.~\ref{fig:gcview}). Left panels: residuals of the parallel component. Right panels: residuals of the perpendicular component. Top: residuals of the relative proper motions. Bottom: residuals of the absolute proper motions. Gray histograms show the full distribution before clipping and blue after removing the $3\sigma$ outliers . Red dotted lines indicate the $3\sigma$ thresholds}
\label{fig:pm_residuals}
    \end{figure}

    For the GNS catalog, we restricted the comparison to stars with proper motion uncertainties smaller than 1.5 mas\,yr$^{-1}$ and cross-matched them with Gaia stars, considering as a positive match any pair of sources within a 50 mas radius. In the top row of Fig.~\ref{fig:pm_residuals}, we show the residuals between Gaia  and GNS. After clipping $3\,\sigma$ outliers, we achieve an rms of $\sim$0.5 mas\,yr$^{-1}$.  The mean offsets in the proper motions parallel and perpendicular to the Galactic Plane are due to the fact that we used a relative frame of reference. The relative reference frame assumes that the mean motion of all stars is zero in all directions. The offset with respect to Gaia corresponds therefore to the relative motion of the Solar System around the GC. This motion is $5.4$\,mas\,yr$^{-1}$ (211\,km\,s$^{-1}$) along the east-west and $0.4$\,mas\,yr$^{-1}$ ($15.6$\,km\,s$^{-1}$) in the north-south direction, in agreement within the uncertainties with what has been measured by radio interferometry \citep[e.g.][]{Reid:2020fl}.

\subsection{Absolute proper motions}\label{sec:absolute_pm}

In the absolute method, proper motions are defined with respect to an absolute reference frame. We used Gaia stars present in the field to anchor our astrometry, following the procedure described below.

First, we selected the most suitable set of Gaia reference stars. We relied on stars from Gaia DR3 \citep{Gaia-Collaboration:2023lr}, which provides a significant improvement in astrometric precision compared to Gaia DR2 \citep{Gaia-Collaboration:2018ux}, reducing the proper motion uncertainty by a factor of $\sim$2 \citep{Gaia_III_precision}. 

Due to the high visual extinction toward the GC, the Gaia mission is largely insensitive to its stellar population, resulting in only a very limited number of usable Gaia stars in this region. However, a few foreground stars have well-measured Gaia astrometry and can serve to link the GNS catalog to the ICRS, an approach previously used in many studies \citep[see, e.g.,][]{Libralato:2021qa, Hosek:2022om, Griggio:2024eb}. Given the small number of available reference stars, careful selection is critical to avoid poor-quality anchors. To this end, we applied the same quality filters described in the preceding section. The remaining stars after these cuts were adopted as the reference stars.

Next, we use Gaia proper motions to propagate their position  to the corresponding GNS epochs. Then, we projected Gaia stars and GNS stars to a common tangential plane and cross-matched Gaia stars with their GNS counterparts, considering sources separated by less than 50\,mas as positive matches. These matched stars were used to compute  similarity transformations, which were applied to the  GNS catalogs to place them into the Gaia reference frame. We cross matched the sources again and refined the alignment by applying a second-degree polynomial transformation. We also tested a first-order polynomial, which left statistically significant positional residuals for many reference stars, and third-order polynomial, which offered no meaningful advantage over the second-degree one. Therefore, we chose a second-order polynomial as the optimal choice.

Finally, we compared the positions of stars common to the GNS\,I\,DR2 and GNS\,II catalogs and divided their positional offsets by the time baseline to derive proper motions. Matches with $H$-band magnitude differences greater than $3\sigma$ were discarded. The resulting proper motions and stellar positions were then compared to Gaia again and the $3\sigma$ outliers were removed of the Gaia reference star list. The  process was repeated until no more 
3$\sigma$ outlier remained. 

As previously, we estimate the uncertainty of the alignment of GNS\,I\, DR2 and GNS\,II with the Gaia reference frame with a bootstrapping method.  In Fig.\ref{fig:B1_align_uncer} bottom panel, we show the alignment uncertainties. Black stars mark the positions of the Gaia reference stars used for the alignment. In Figure~\ref{fig:pm_residuals} bottom panel, we  show the residuals of the absolute proper motions with Gaia. As in the case of the relative proper motions, we reach an rms of $\sim$0.5\,mas\,yr$^{-1}$. Both methodologies present similar proper motion residuals with respect to Gaia, confirming the consistency and robustness of both approaches. 

In the bottom panel of Fig.~\ref{fig:B1_align_uncer}, we observe an almost homogeneous distribution of the alignment error across the field. However, a slight dependence of the alignment quality on the local Gaia stellar density is noticeable. As we will see in the next section, this effect becomes more pronounced in the Arches field (Fig.~\ref{fig:Arches_align_uncer}, right panel), where the number of Gaia stars is lower and their spatial distribution is highly heterogeneous. Consequently, alignment uncertainties are larger in regions with fewer available Gaia reference stars. This contrasts with the case of the relative alignment, where the effective density of reference stars is high and largely homogenous  across the field, resulting in a more uniform uncertainty distribution.


\subsection{NSD field}

As a secondary test, we analysed a field containing the Arches cluster (blue solid box in Fig. \ref{fig:gcview}) to assess the quality of the proper motions in a highly crowded region, compare the measured proper motions to literature values, and study the feasibility of detecting clusters in such an environment.

\begin{figure}[h!]
    \centering
    \includegraphics[width=0.49\linewidth]{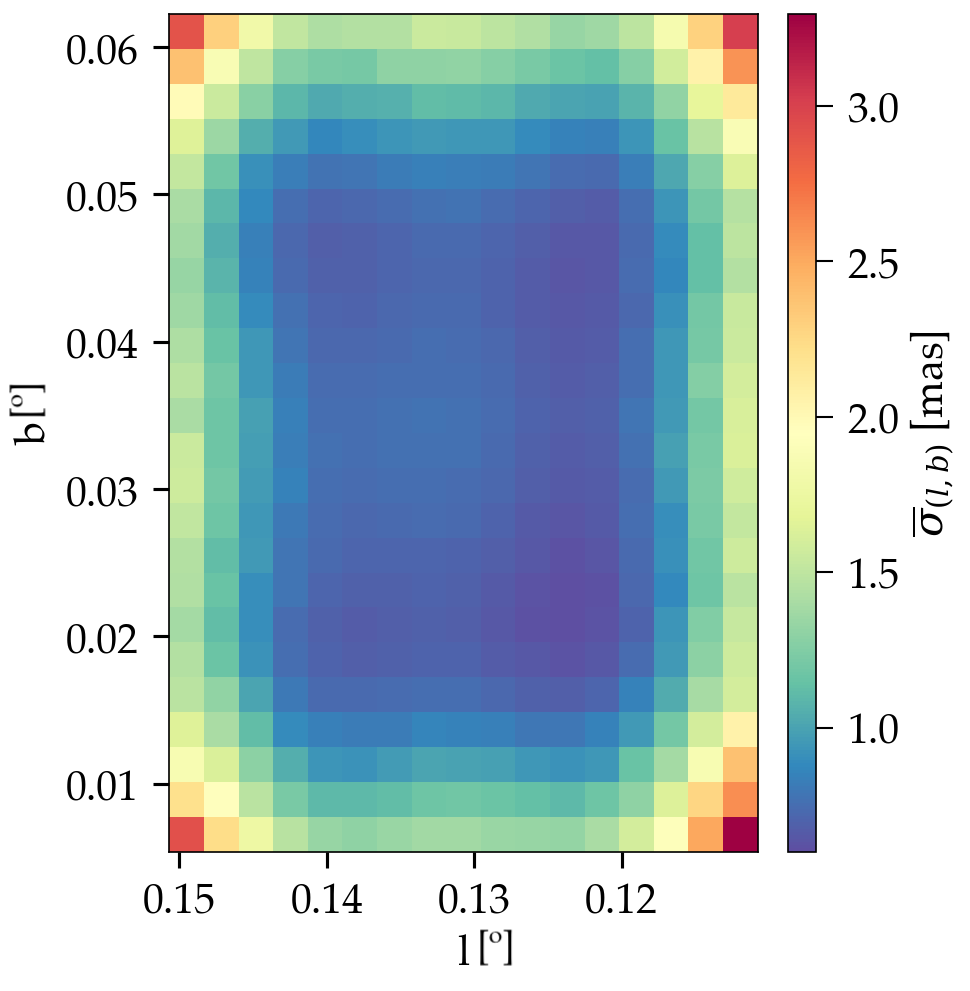}  
    \includegraphics[width=0.49\linewidth]{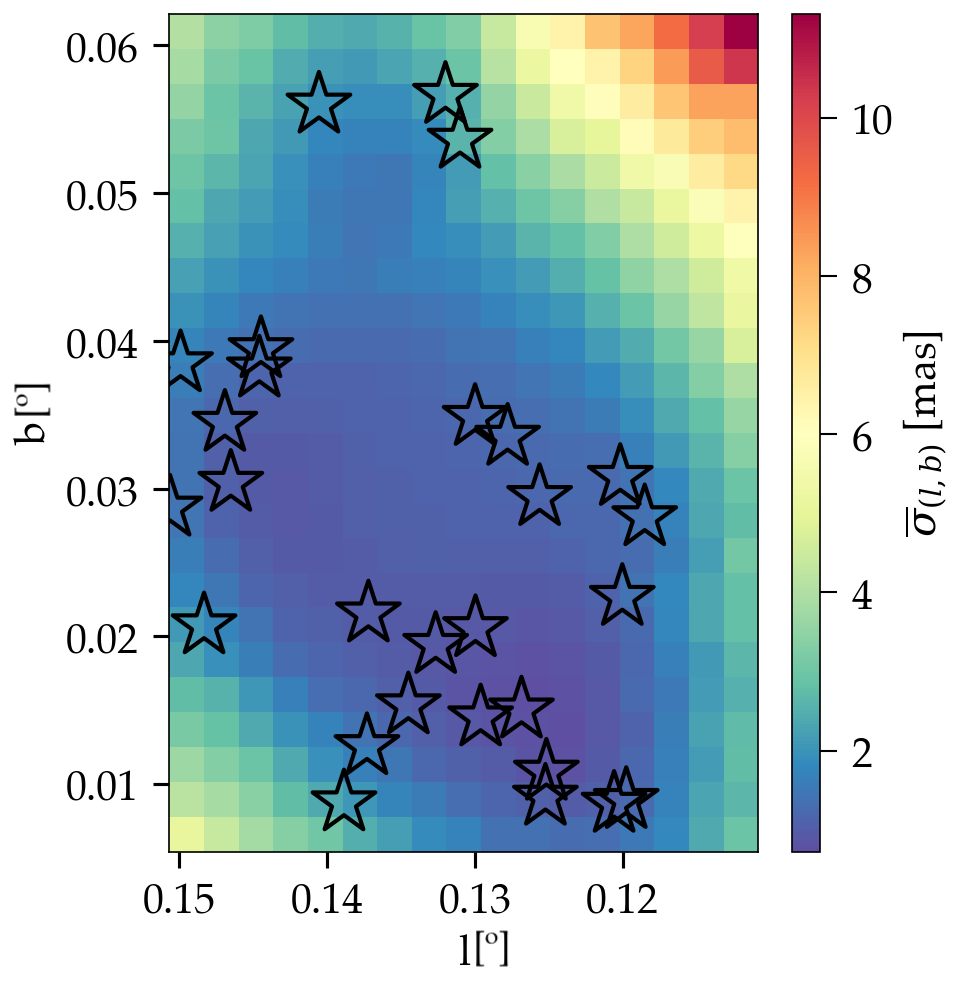}  
    \caption{Alignment uncertainty for the solid blue area in Fig. \ref{fig:gcview}. Left: Relative alignment. Right: Absolute alignment. Black stars mark the position of Gaia reference stars.}
    \label{fig:Arches_align_uncer}
\end{figure}

\begin{figure}[h!]
    \centering
    \includegraphics[width=1\linewidth]{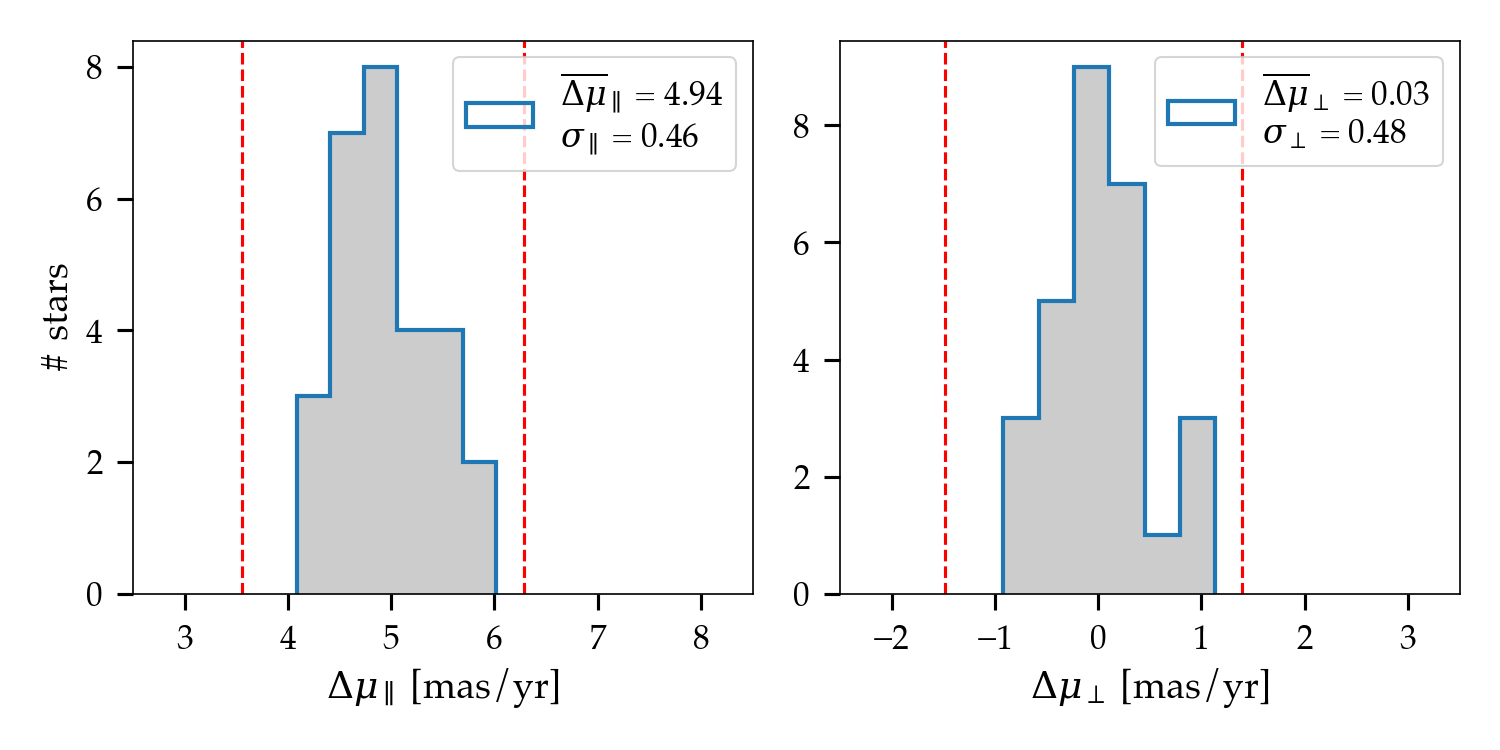}\\[1ex] 
    \includegraphics[width=1\linewidth]{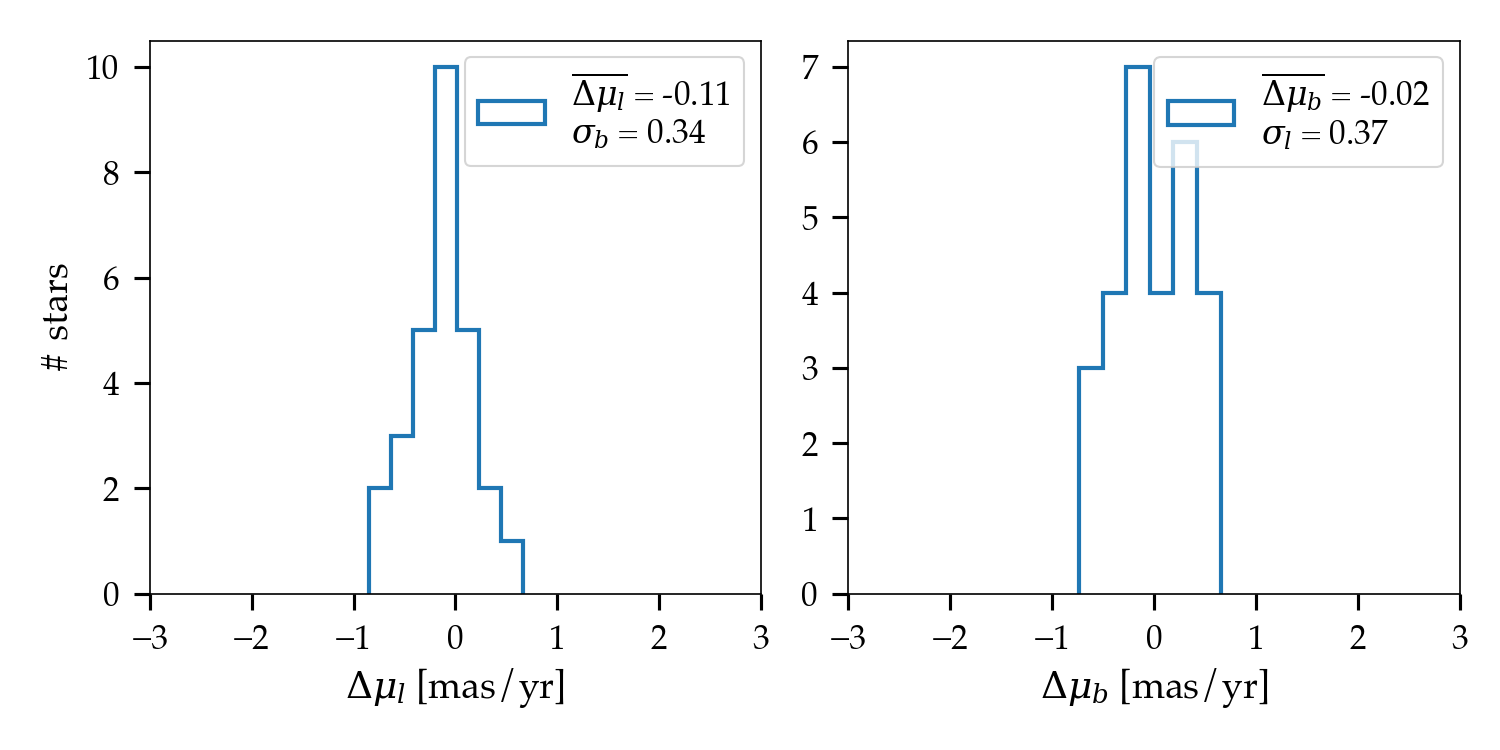}       
    \caption{Gaia–GNS proper motion residuals Arches field in Fig.\ref{fig:gcview}. Top: relative proper motions residuals  Bottom: absolute proper motions residuals. Gray histograms show the full distribution before clipping. Red dotted lines indicate the $3\sigma$ thresholds}
    \label{fig:pm_Arches_residuals}
    \end{figure}

We compute both relative and absolute proper motions, as described in Sections~\ref{sec:relative_pm} and \ref{sec:absolute_pm}. The alignment uncertainties for the region containing the Arches cluster are shown in Fig.~\ref{fig:Arches_align_uncer}. The left panel displays the results for the relative uncertainties, which, as in the case of the test field on the Galactic bar, show mostly homogeneous values. The right panel presents the alignment uncertainties of the absolute method together with the Gaia reference stars. A strong dependence of the uncertainty on the density of Gaia stars in the field is evident. Figure~\ref{fig:pm_Arches_residuals} shows the proper-motion residuals for both the relative and absolute measurements.

We also tested the quality the relative and absolute proper motions by comparing them with the catalogue of \citet{Hosek:2022om}, hereafter H22. This catalogue consists of astro-photometry data, proper motions, and magnitudes (F127M and F153M filters) obtained with the WFC3/HST camera in the area of the Arches cluster (black box in Fig.~\ref{fig:gcview}). The precision of the H22 proper motions reaches $\sim$0.2\,mas\,yr$^{-1}$ rms when compared to Gaia DR3 and, like our absolute proper motions, the H22 catalogue is anchored to the ICRS using Gaia DR3 stars.

For the comparison between the GNS and H22 catalogues, we selected stars with low position uncertainties,i.e.,  for GNS, stars with $12 < H < 18$, and for H22, stars with $15 < F153M < 20$ (see Fig.~2 in \citealt{Hosek:2022om}). In both cases we discarded stars with proper motion uncertainties larger than 0.5\,mas\,yr$^{-1}$.  

\begin{figure}[h!]
    \centering
    \includegraphics[width=1\linewidth]{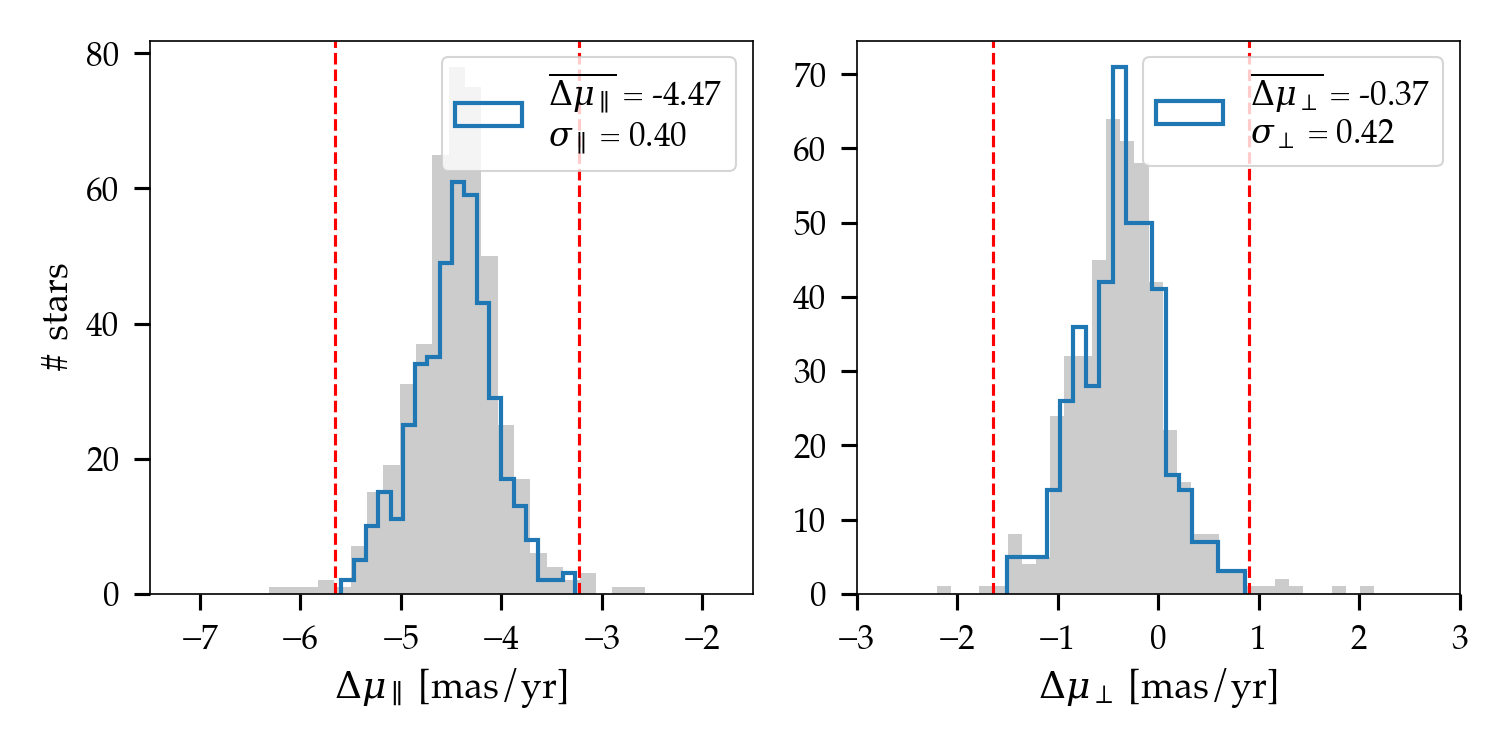}\\[1ex] 
    \includegraphics[width=1\linewidth]{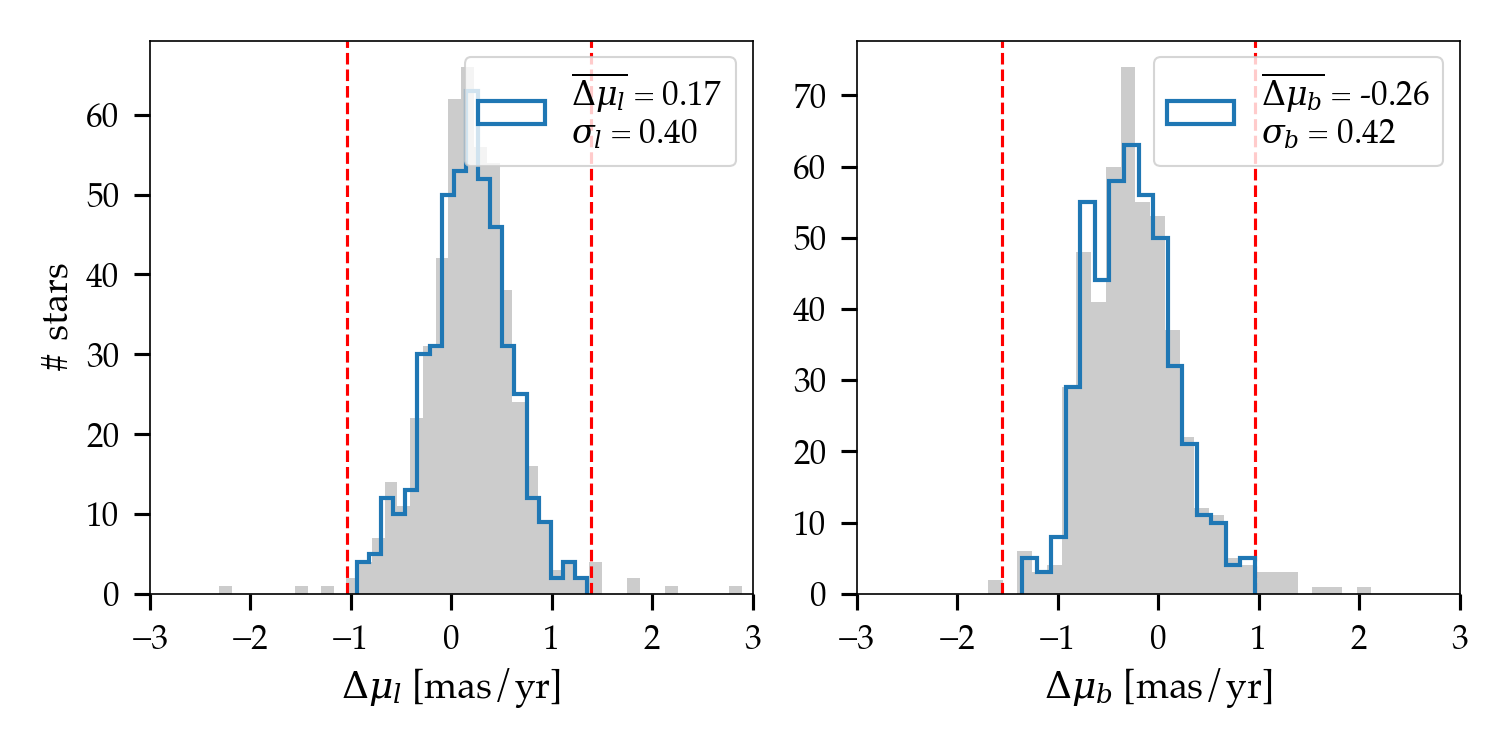}       
   \caption{Residuals between H22 and GNS proper motions (blue solid field and black square field in Fig.~\ref{fig:gcview}). Top: residuals of the relative proper motions. Bottom: residuals of the absolute proper motions. Gray histograms show the full distribution before clipping. Red dotted lines indicate the $3\sigma$ thresholds}
    \label{fig:pm_Hosek_residuals}
    \end{figure}

    Figure~\ref{fig:pm_Hosek_residuals} shows the residuals of the comparison between H22 and our relative proper motions (top panel), and between H22 and our absolute proper motions (botton panel). We achieve a precision of $\sim 0.4 \,\mathrm{mas\,yr^{-1}}$ rms in both cases. These low and homogeneous residuals across both methodologies further demonstrate the reliability and compatibility of the two methods.

Finally, we applied the cluster-finding algorithm described in \citet{Martinez-Arranz:2024gm} to both the relative and absolute proper motion catalogues overlapping with the Arches cluster. This tool is based on the Density-Based Spatial Clustering of Applications with Noise algorithm \citep[DBSCAN;][]{Ester:1996}, which identifies overdensities in multidimensional spaces. In this case, we searched for overdensities in a four-dimensional space defined by celestial positions and proper motions.

\begin{figure}[h!]
    \centering
    \includegraphics[width=1\linewidth]{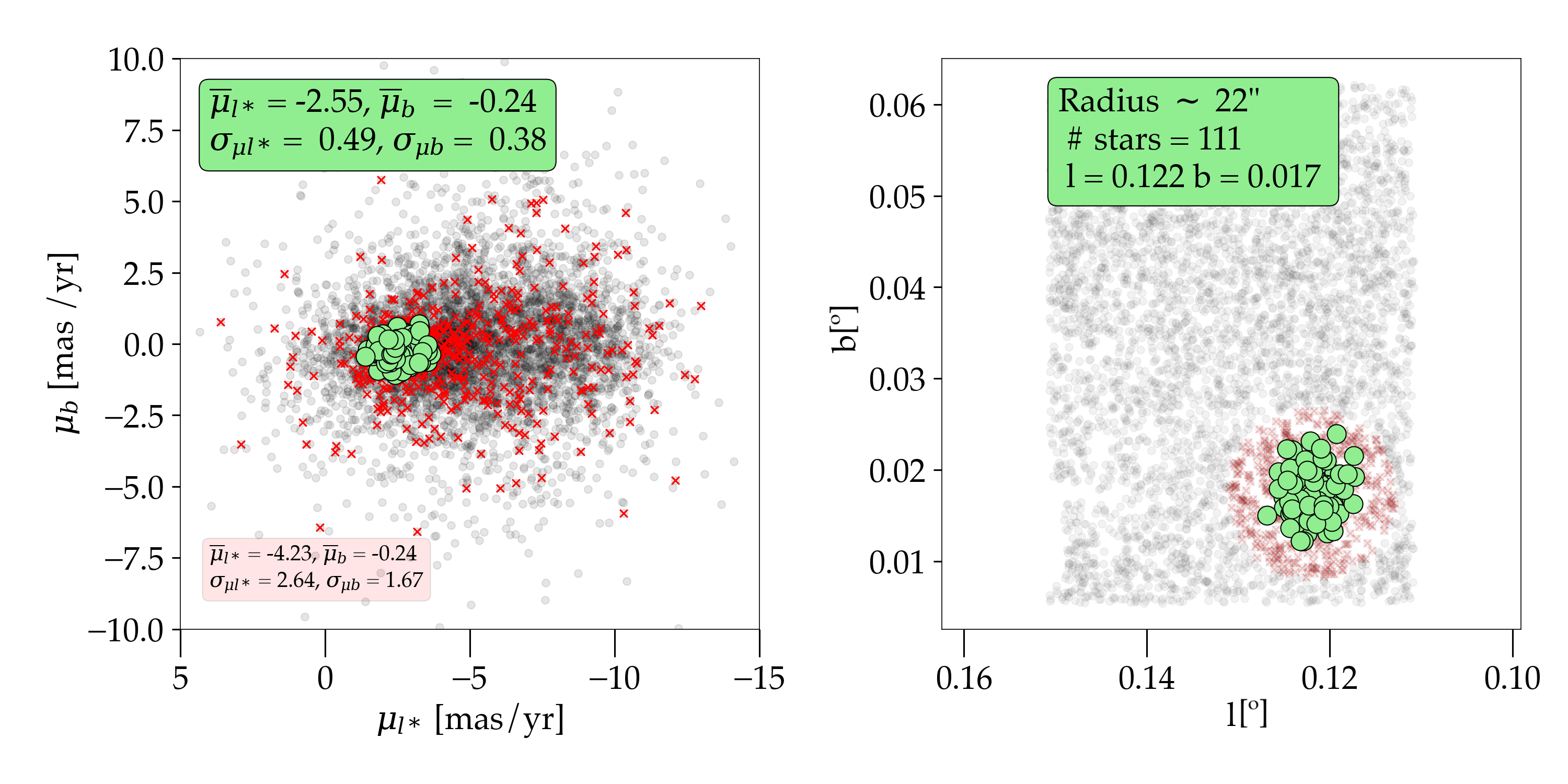}\\[1ex] 
    \includegraphics[width=1\linewidth]{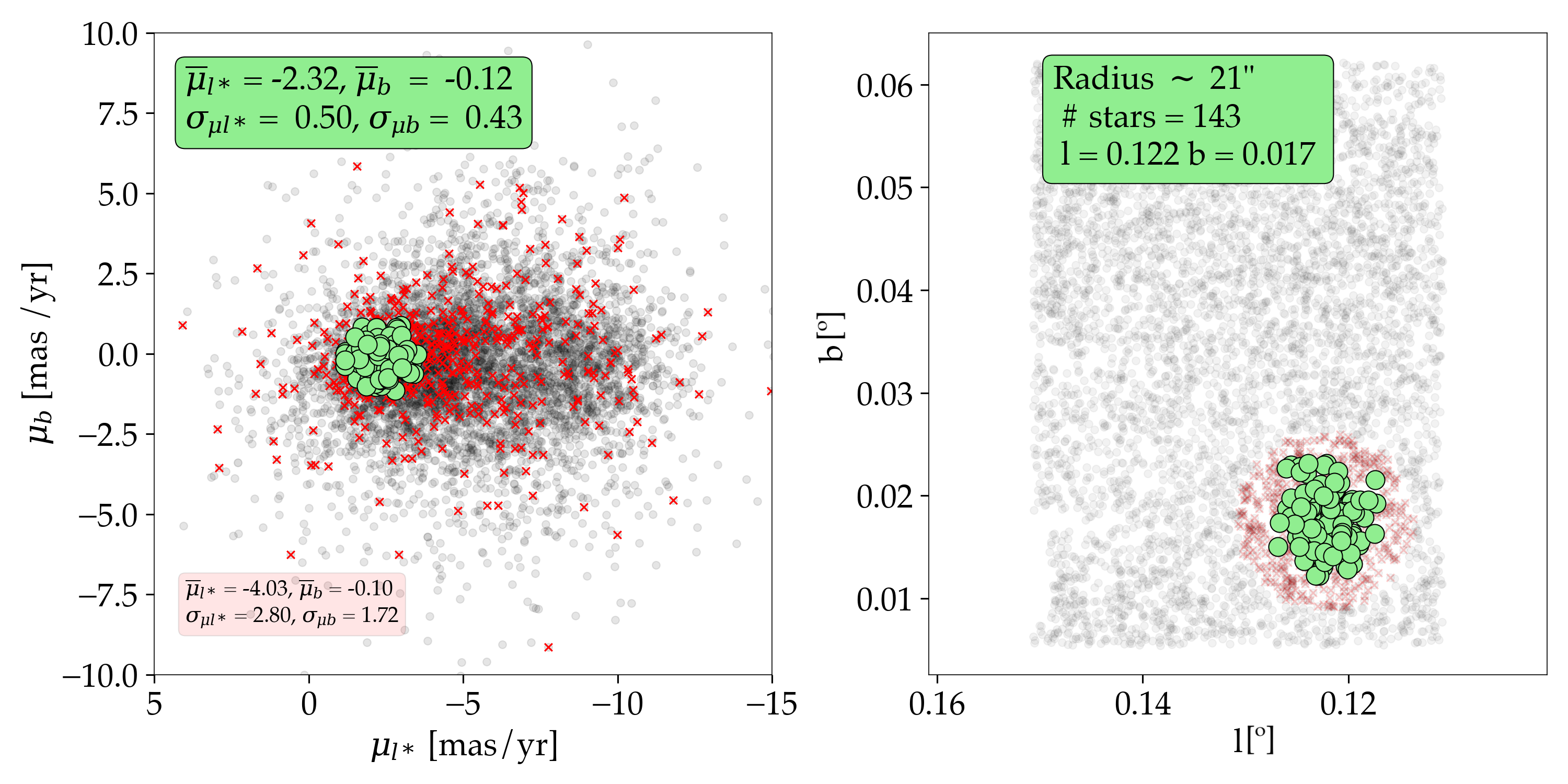}
   \caption{Comoving groups identified in the Arches field (blue box in Fig.~\ref{fig:gcview}). The left column shows proper motions, while the right column shows positions. The top left panel displays relative proper motions, after subtracting the mean residuals with respect to Gaia. The bottom left panel displays absolute proper motions. Green points represent the stars identified by the clustering algorithm as members of a comoving group, red crosses mark stars within 1.5 times the radius of the comoving group, and black points correspond to field stars. The boxes indicate the mean proper motion values with their dispersions, the approximate radius of the comoving group, the number of identified members, and their mean positions.}

    \label{fig:Arches_gns_cluster}
    \end{figure}

For the relative proper motions, we subtracted the mean residuals with respect to Gaia in order to place them in the Gaia reference frame. The analysis was restricted to stars with proper motion uncertainties below 1.5\,mas\,yr$^{-1}$. In both cases, we identified a dense co-moving group of stars. Figure~\ref{fig:Arches_gns_cluster} shows the co-moving group detected in the relative and absolute catalogues. On the left column, we display a vector-point diagram of the proper motions for each catalog, relative at the top and absolute at the bottom. On the right column, we show the star positions from each catalog. The green dots represent the stars labeled as members of the same co-moving group by the clustering algorithm. 

The members of the co-moving groups identified in both catalogs, relative and absolute, overlap with the known extent of the Arches cluster. Both groups exhibit consistent mean velocities, within the uncertainties, as well as coincident spatial distributions. The mean velocities parallel and perpendicular to the Galactic plane agree in both cases with the values reported for the Arches cluster by \citet{Hosek:2022om} ($\mu_{l*} = -2.03 \pm 0.025$\,mas\,yr$^{-1}$, $\mu_{b} = -0.30 \pm 0.029$\,mas\,yr$^{-1}$) and \citet{Libralato:2021qa} ($\mu_{l*} = -3.05 \pm 0.17$\,mas\,yr$^{-1}$, $\mu_{b} = -0.16 \pm 0.20$\,mas\,yr$^{-1}$). A more detailed analysis of the co-moving groups identified in our data sets lies beyond the scope of this paper and will be presented in a forthcoming work (Martínez-Arranz et al., in prep.).

\section{Conclusions}

We have described the updated methodology applied to produce data release 2 (DR2) of the GALACTICNUCLEUS (GNS) survey. DR2 builds upon the first epoch of $JHK_s$ imaging presented by \citet{Nogueras-Lara:2019yj}. In addition, we include the $H$-band observations from a second epoch of GNS imaging. In this work, we focus on detailing the new reduction, calibration, and analysis procedures, and we demonstrate their performance using two representative test fields. The full catalogue will be released in a future publication.

The updated data reduction and analysis methods allow us to reach five milli arcseconds absolute astrometry, an improvement of a factor $\sim$10 compared to GNS\,I DR\,1. The photometric uncertainties are generally $<5\%$ in all bands. Their basic limitation is the $\sim$3\% zero point uncertainty of the SIRIUS/IRSF GC survey, that we use for photometric calibration \citep{Nagayama:2003fk,Nishiyama:2006ai}. Finally, the GNS\,I DR\,2 is about 0.5\,mag deeper than DR\,1, which is mostly due to the adoption of a jackknife algorithm to estimate the uncertainties.

We present the methodology and performance of the first proper motion measurements based on two epochs of GNS $H$-band imaging. Despite relying solely on ground-based observations and only two epochs separated by seven years (six in the case of the fields~B1 and~20) , the proper motions achieve a precision comparable to that obtained with space-based, multi-epoch data. We implemented and compared two independent methods to define the reference frame and obtained consistent results. This dual approach provides flexibility to select the most suitable methodology depending on the characteristics of each field, such as stellar density, extinction, or the availability of Gaia reference stars.

We demonstrated that alignment with the Gaia reference frame is not always optimal, because this strongly depends on the availability and spatial distribution of Gaia stars in a given field. In some cases, relative proper motions provide a robust alternative. Conversely, in regions of very high extinction where relative alignment becomes less effective, the absolute method based on Gaia stars can still be applied successfully.

Our proper motions achieves an accuracy comparable to that of space-based studies. This confirms that the technique can be further extended by incorporating data from  space-based telescopes such as JWST or the future Roman Space Telescope. Such extensions will allow us to significantly improve on the current $\sim$0.5\,mas\,yr$^{-1}$ accuracy (rms with comparison to Gaia) that we can currently achieve.

Two of the test files we reduced overlap with the Arches cluster. The clustering algorithm described by \citet{Martinez-Arranz:2024gm} successfully identified the Arches cluster, with mean velocity components consistent with previous studies based on fundamentally different data sets and methodologies \citep{Hosek:2022om,Libralato:2020}. This highlights the effectiveness of our approach for detecting and characterizing stellar clusters in the crowded and complex environment of the GC.

As demonstrated by \citet{Martinez-Arranz:2024gm} and \citet{Martinez-Arranz:2024nr} for the case of the Candela\,1 cluster, the future GNS proper motion catalogue will enable us to search for and characterize so far unknown stellar associations and streams in regions suspected to host recent star formation, such as Sgr~B1 and Sgr~C (see Fig.~\ref{fig:gcview}), where recent studies point to the presence of $\sim10^{5}$\,M$_\odot$ of newly formed stars \citep{Nogueras-Lara:2022jz,Nogueras-Lara:2024gf}. This opens the possibility of studying the kinematics of the youngest stellar populations in the NSD, thereby addressing questions such as the missing cluster problem or testing whether the IMF is fundamentally different in this extreme environment compared to the Galactic disk, as some studies have suggested \citep{Morris:1993ve, Bartko:2010fk,Hosek:2019vn}. The GNS proper motions catalogue will also provide us with a new tool to understand the structure and formation history of the NSD. Work is currently underway on the data release of these catalogues, which will make such studies possible in the near future.

\begin{acknowledgements}
 AMA, RS, and FNL acknowledge financial support from the Severo Ochoa
 grant CEX2021-001131-S funded by MCIN/AEI/ 10.13039/501100011033  and from grant
 PID2022-136640NB-C21 funded by MCIN/AEI 10.13039/501100011033 and by
 the European Union, as well as support from grant PID2024-162148NA-I00, funded by MCIN/AEI/10.13039/501100011033 and the European Regional Development Fund (ERDF) “A way of making Europe”.
\end{acknowledgements}

%
  \bibliographystyle{aa} 
   \bibliography{BibGC} 

\end{document}